\documentclass[twocolumn, twocolappendix]{aastex631}

\usepackage{newtxtext,newtxmath,xcolor}
\usepackage{amsmath, graphics, bm, graphicx, wasysym, verbatim}
\allowdisplaybreaks

\newcommand{\mathsym}[1]{{}}
\newcommand{\unicode}[1]{{}}

\newcommand{\cg}{\gamma}

\newcommand{\dg}{\delta}
\newcommand{\Dg}{\Delta}

\newcommand{\Om}{\Omega}
\newcommand{\om}{\omega}

\newcommand{\pd}{\partial}
\newcommand{\im}{{\rm i}}
\newcommand{\der}{{\rm d}}

\newcommand{\eps}{\varepsilon}

\newcommand{\bxi}{{\bm \xi}}

\newcommand{\e}{{\rm e}}

\newcommand{\hn}{{\bm {\hat n}}}

\newcommand{\he}{{\bm {\hat e}}}

\newcommand{\vphi}{\varphi}

\newcommand{\del}{\nabla}
\newcommand{\br}{{\bm r}}

\newcommand{\Mp}{M_{\rm p}}

\newcommand{\bcdot}{{\bm \cdot}}
\newcommand{\bdel}{{\bm \del}}

\newcommand{\Rp}{R_{\rm p}}

\newcommand{\be}{\begin{equation}}
\newcommand{\ee}{\end{equation}}

\begin{document}

\title{Seismic Oscillations Excited by Giant Impacts in Directly-Imaged Giant Planets}

\author[0000-0002-9849-5886]{J. J. Zanazzi}\thanks{51 Pegasi b fellow, email:jzanazzi@berkeley.edu}
\affiliation{
Astronomy Department, Theoretical Astrophysics Center, and Center for Integrative Planetary Science, University of California, Berkeley, \\
Berkeley, CA 94720, USA \\
}

\author[0000-0002-6246-2310]{Eugene Chiang}
\affiliation{
Astronomy Department, Theoretical Astrophysics Center, and Center for Integrative Planetary Science, University of California, Berkeley, \\
Berkeley, CA 94720, USA \\
}
\affiliation{
Department of Earth and Planetary Science, University of California, Berkeley, CA 94720, USA
}

\author[0000-0003-2969-6040]{Yifan Zhou}
\affiliation{
Department of Astronomy,
University of Virginia, 
530 McCormick Rd.,
Charlottesville, VA 22904, USA
}




\begin{abstract}
    The radii and masses of many giant exoplanets imply their interiors each  contain more than $\sim$100 $M_\oplus$ of solids. A large metal content may arise when a giant planet grows by colliding and merging with multiple $\sim$10 $M_\oplus$ solid cores.  Here we show that a giant impact with a young gas giant excites long-lived seismic oscillations that can be detected photometrically.  Mode lifetimes are close to the planet's Kelvin-Helmholtz time, a significant fraction of a young planet's age.  Oscillation periods lie between tens of minutes to an hour, and variability amplitudes can exceed a percent for several million years.  Beta Pictoris b is a young super-Jupiter known to be highly metal-enriched. If a Neptune-mass (17 $M_\oplus$) body impacted $\beta$ Pictoris b in the past $\sim$9--18 Myr, the planet could still be ringing with a percent-level photometric variability measurable with JWST.
\end{abstract}



\section{Introduction}

Extrasolar gas giants are often metal-enriched.  
Measurements of a giant planet's mass and radius can be used to infer its heavy-metal mass fraction $Z_{\rm p}$. \cite{Thorngren+(2016)} has shown $Z_{\rm p}$ often exceeds the metallicity of the host star $Z_\star$ by large factors. 
Giant planet heavy metal masses can exceed $\sim$100 $M_\oplus$, much larger than the critical core masses of $\sim$5-20 $M_\oplus$ needed to nucleate giant planet formation.  Later works have found similar correlations in the composition of gas giants with the planet's mass, using either constraints on the planet's bulk density, or atmospheric retrievals \citep[e.g.][]{ThorngrenFortney(2018), Teske+(2019), Welbanks+(2019), Muller+(2020), Nikolov+(2022), Bean+(2023), Sun+(2024), Wang2025}.  \cite{Thorngren+(2016)} posited this correlation arose from the preferential accretion of dust grains or planetesimals drifting through gas in the protoplanetary disk (see also e.g. \citealt{Hasegawa+(2018), Hasegawa+(2019), ShibataIkoma(2019), Shibata+(2020), Venturini+(2020), SchneiderBitsch(2021), Knierim+(2022), Morbidelli+(2023), Danti+(2023), BitschMah(2023)}).  Alternatively, \cite{GinzburgChiang(2020)} showed that collisions between multiple protoplanet cores (`major mergers'), each undergoing runaway gas accretion, would enrich the resultant gas giant in metals and produce a $Z_{\rm p}/Z_\star$ ratio that decreases with $M_{\rm p}$, as observed (see also \citealt{DawsonMurrayClay13}; \citealt{Frelikh+2019}; \citealt{Ogihara+(2021)}; \citealt{Li+2021}).  These two theories give competing visions for how giant planets acquire their metallicities: the first by quiescently accreting disk solids which slip into the gas gap opened by the planet, and the other by violently engulfing nearby protoplanets.

Solar System giants experience collisions large and small. The large obliquity of Uranus, as well as the planet's rings and regular moons, could be the result of a giant impact \citep[e.g.][]{Kegerreis+(2018), Reinhardt+(2020)}. It has been proposed that the dilute core of Jupiter arose after an impact with a $10 M_\oplus$ body 
(\citealt{Liu+(2019)}, but see also \citealt{Sandes+2024, Meier+2025}). On a much smaller scale, the comet Shoemaker-Levy 9 (SL9) struck Jupiter in 1994. 
Several papers pointed out the SL9 impact would excite a spectrum of normal mode oscillations \citep{Kanamori(1993), Marley(1994), LeeVanHorn(1994), Lognonne+(1994)}. The resultant atmospheric temperature variations proved too small to be detected in the mid-infrared by the 3.6-meter telescope at the European Southern Observatory \citep{DombardBoughn(1995), Mosser+(1996)}.  Seismic oscillations in Jupiter (too large to be excited by SL9) were reported to have been detected in 2011 by a team at the Nice Observatory, using radial velocity measurements gathered by the SYMPA Fourier spectro-imager \citep{Schmider+(2007), Gaulme+(2008), Gaulme+(2011)}, although follow-up Doppler measurements by \cite{Shaw+2022} have failed to confirm their detection. Saturn's normal modes excite density waves within the planet's rings \citep[e.g.][]{Marley(1991), MarleyPorco(1993), Marley(2014)}.  A number of works \citep{HedmanNicholson(2013),HedmanNicholson(2014), French+(2016), French+(2019), Hedman+(2019)} used Cassini data on these waves as a seismograph to measure the oscillation amplitudes and frequencies of Saturn's fundamental modes, inferring stable stratification deep within Saturn's interior \citep{Fuller(2014), MankovichFuller(2021)}.   Residual data from the Cassini gravity experiment could also be indicative of Saturn pressure modes \citep{Markham+2020}.  A mystery for both Jupiter and Saturn is how their oscillations are excited. \cite{MarkhamStevenson(2018)} favor rock storms deep in Jupiter's atmosphere, while \cite{WuLithwick(2019)} advocate collisions to excite the spectrum of modes seen in Saturn.

In principle, planet-scale impacts could excite seismic oscillations in directly imaged exoplanets, which could be detected by space-based missions such as JWST and Roman.  Directly imaged exoplanets tend to be young ($\lesssim 20$ Myr), massive ($\gtrsim 10 M_{\rm Jup}$), bright, and embedded in either protoplanetary disks \citep[PDS 70 bc,][]{Keppler+(2018), Haffert+(2019), Wang+(2020)} or debris disks ($\beta$ Pic b, e.g. \citealt{Lagrange+(2009), Currie+(2013), Wang+(2016), Lagrange+(2019)};  and HR 8799 bcde, \citealt{Marois+(2008), GRAVITY(2019), Faramaz+(2021)}).  Such planets could have undergone major mergers in the past few million years and had their normal modes excited to observable amplitude.


In this paper, we take as a test case $\beta$ Pictoris b ($\beta$ Pic b).  The $13 M_{\rm Jup}$ planet and its birth environment suggest it may have suffered multiple mergers with other massive protoplanets.  Atmospheric retrievals indicate the planet's heavy-metal mass lies between $\sim$100-300 Earth masses \citep[e.g.][]{GRAVITY2020, Wang2025}.  The planet's eccentricity of $\sim$0.15 \citep{GRAVITY2020}  might have been excited by an impact with a body originating from the system's dynamically active debris disk \citep[e.g.][]{Dawson+(2011), Rebollido2024}.  The young age of the $\beta$ Pic moving group ($\sim$20 Myr, e.g. \citealt{MamajekBell2014}) implies that if an impact occurred, normal modes would not have much time to damp.  And given the planet's luminosity and large separation from its host star, JWST photometry might be able to detect seismic oscillations. Section~\ref{sec:osc_review} reviews the physics behind giant planet $f$-mode and $p$-mode oscillations.  Section~\ref{sec:ModeAmp} calculates the amplitudes of normal modes excited by a merger.  Section~\ref{sec:ModeDamp} shows that mode damping timescales from internal friction are of order the Kelvin-Helmholtz time, so oscillations can survive a significant fraction of the planet's age.  Section~\ref{sec:LumVar} shows JWST observations might detect the luminosity variations from these seismic oscillations.  Section~\ref{sec:SummDisc} provides a summary, and discusses future avenues of investigation for exoplanet systems.

\section{Giant Planet Oscillation Modes}
\label{sec:osc_review}

\begin{figure*}
\centering
\includegraphics[width=0.75\linewidth]{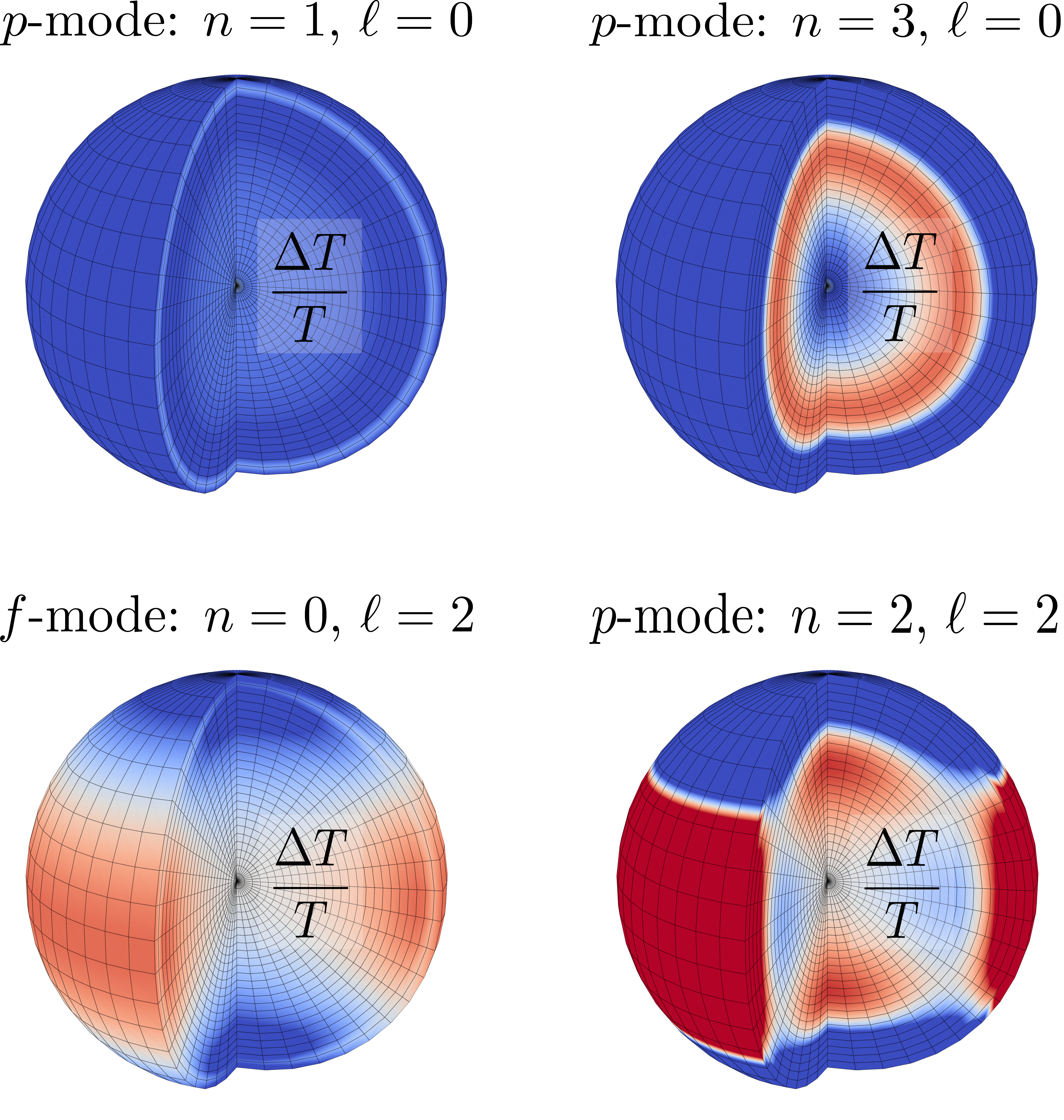}
\caption{Visualizations of $f$-modes ($n=0$) and $p$-modes ($n>0$), for the angular degrees $\ell$ indicated.  Interior colors display a temperature excess ($\Delta T > 0$, red) or deficit ($\Delta T < 0$, blue) due to the oscillation.  The color scale is linear in all diagrams, although the maximum and minimum $\Delta T/T$ values vary, clipping the color scale to retain detail closer to the planet's center.
\label{fig:ModeVis}}
\end{figure*}

\begin{figure}
\includegraphics[width=\linewidth]{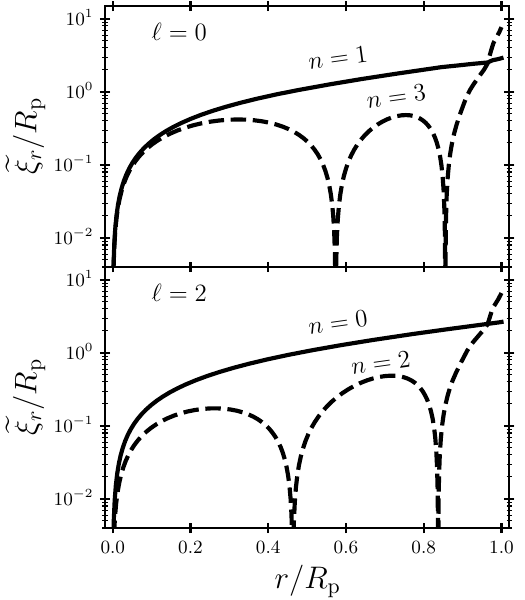}
\caption{We display $\widetilde \xi_r$ profiles for $f$-modes ($n=0$) and $p$-modes ($n>0$), for the angular degrees $\ell$ and radial node numbers $n$ indicated.  The number of radial nodes $n$ equals the number of times $\mathcal{Y}_1$ crosses zero in the radial direction (eq.~\ref{eq:Y_1}), excepting the origin.  When the mode's self-gravity is small ($\dg \widetilde \Phi \approx 0$), $n$ counts the nodes in the radial displacement $\widetilde \xi_r$.  We normalize $\widetilde \xi_r$ using equation~\eqref{eq:norm}.
\label{fig:ModeProfs}}
\end{figure}

\begin{figure}
\includegraphics[width=\linewidth]{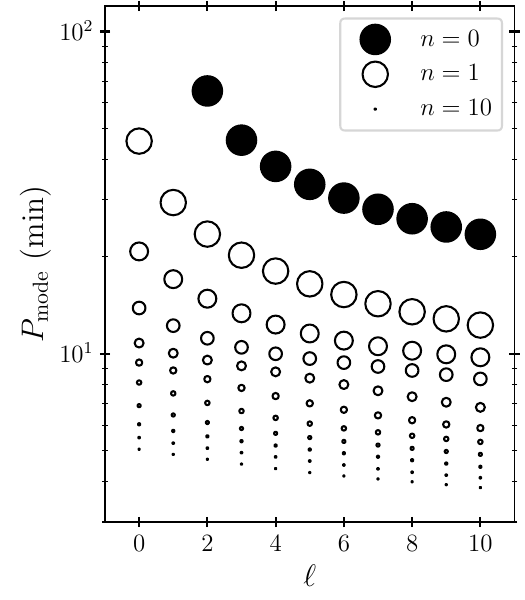}
\caption{Mode periods $P_{\rm mode} = 2\pi/\omega$, vs.~angular degree $\ell$ and radial node number $n$.  The $f$-modes are displayed by large filled circles, while $p$-modes are open circles whose size decreases as $n$ increases from $n=1$ to $n=10$.  Notice $\ell < 2$ do not have $f$-modes.  Mode periods $P_{\rm mode}$ decreases as $\ell$ and $n$ increase.  
\label{fig:ModeP}}
\end{figure}

Giant planets support both fundamental ($f$-mode) and pressure ($p$-mode) oscillations, for which the main restoring forces are gravity and pressure. The Lagrangian perturbation to the temperature for each $f$-mode and $p$-mode can be decomposed into spherical harmonics $\Delta T/T \propto Y_{\ell m}(\theta, \vphi)$.  When 
the oscillation is axisymmetric (azimuthal number $m=0$), the angular degree $\ell$ equals the number of latitudinal ($\theta$) nodes in $\Delta T/T$ (Fig.~\ref{fig:ModeVis}).  For a given $\ell$, the mode's node number $n$ counts the radial nodes in the function
\be
\mathcal{Y}_1 = \left(1 - \frac{\rho}{\rho_{\rm av}}\right) \frac{\xi_r}{r} + \frac{1}{3g} \left( \frac{\dg \Phi}{r} - \frac{\der \dg \Phi}{\der r} \right),
\label{eq:Y_1}
\ee
excepting the origin, where $\dg \Phi$ is the Eulerian perturbation to the gravitational potential, $g = G M_r/r^2$ is the gravitational acceleration, $M_r = 4\pi \int_0^r \rho r^2 \der r$ is total mass enclosed within radius $r$, and $\rho_{\rm av} = \frac{3 M_r}{4\pi r^3}$ \citep[see][for details]{Takata2006}.  The $f$-mode has $n = 0$ nodes in $\mathcal{Y}_1$ (and the radial displacement $\xi_r$),
and has the lowest frequency $\omega$, while all $p$-modes have $n>0$, and frequencies $\omega$ that increase with $n$ (Figs.~\ref{fig:ModeProfs} and~\ref{fig:ModeP}).  Fundamental modes do not exist for $\ell = 0,1$.
In all calculations below, we use a \texttt{MESA} model for a super-Jupiter similar to $\beta$ Pic b, with mass $M_{\rm p} = 13 \ M_{\rm Jup}$, radius $R_{\rm p} = 1.4 \ R_{\rm Jup}$, effective temperature $T_{\rm eff} = 1720 \ {\rm K}$, luminosity $L_{\rm p} = 1.6 \times 10^{-4} \ L_\odot$, and age $18 \ {\rm Myr}$, and calculate  oscillations using the \texttt{GYRE} code, with inlists given in Appendix~\ref{app:inlist}.  For simplicity, we ignore how a solid core affects the oscillation spectrum, which could alter the $f$-mode and $p$-mode frequencies by several percent \citep[e.g.][]{LiBihanBurrows(2013), Fuller+(2014)}.

\begin{figure}
\includegraphics[width=\linewidth]{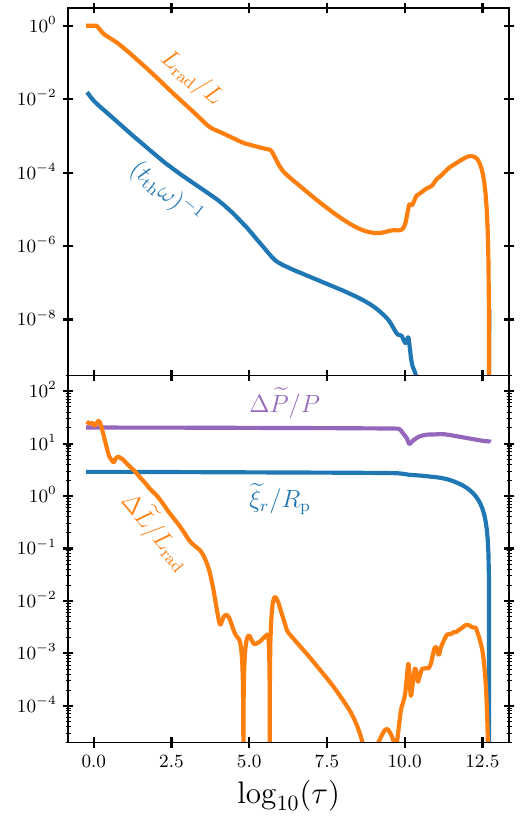}
\caption{Profiles of different quantities with optical depth $\tau$ inside the super-Jupiter. 
 \textit{Top panel}: Ratio of radiative to total internal luminosity $L_{\rm rad}/L$, and thermal time $t_{\rm th}$ (eq.~\ref{eq:t_th}) multiplied by the mode frequency $\omega$, for the $(\ell, n)=(0,1)$ $p$-mode.  Energy transport by radiation dominates only near the photosphere ($\tau \lesssim 1$), and oscillations are approximately adiabatic everywhere ($t_{\rm th} \omega \gg 1$).  \textit{Bottom panel}:  Comparing how the mode perturbs the radiative luminosity 
 $\Delta \widetilde L/L_{\rm rad}$, 
 compared to the radial displacement $\widetilde \xi_r/R_{\rm p}$ and Lagrangian pressure perturbation $\Dg \widetilde P/P$.  
 Jumps in $\Delta \widetilde L/L_{\rm rad}$ at large optical depths ($\tau \gg 1$) occur because of variations in the opacity. 
The transition of hydrogen from ionized to neutral near $\tau \sim 10^{10}$ affects the adiabatic sound speed, causing the small dip in $\Delta \widetilde P/P$.
\label{fig:FModeProf}}
\end{figure}

We define $t_{\rm th}$ as the thermal time of a shell of thickness equal to the local pressure scale height $H_P = -(\der \ln P/\der r)^{-1}$:
\begin{equation}
    t_{\rm th} = \frac{4\pi r^2 H_P \rho C_P T}{L},
    \label{eq:t_th}
\end{equation}
where $\rho$ is the density, $C_P$ the specific heat at constant pressure, $T$ the temperature, 
and $L$ the local luminosity 
(radiative and convective). 
Comparing $t_{\rm th}$ to the oscillation frequency $\omega$ of the $(\ell, n)=(0, 1)$ $p$-mode in Fig.~\ref{fig:FModeProf} (top panel), we see $t_{\rm th} \omega \gg 1$. 
We conclude that this mode is approximately adiabatic everywhere (e.g.~\citealt{Unno+(1989), Pfahl+(2008)}).
Since all $f$ and $p$-modes have frequencies comparable to or larger than the $(\ell,n)=(0,1)$ $p$-mode (Fig.~\ref{fig:ModeP}), these other oscillations are nearly adiabatic as well.

The luminosity emergent from the surface of the planet is carried by radiation (Fig.~\ref{fig:FModeProf}, top panel). 
Both $f$-mode and $p$-mode oscillations perturb the radiative luminosity 
\begin{equation}
    L_{\rm rad} = - \frac{4\pi r^2 c}{3 \kappa \rho} \frac{\der}{\der r}(a T^4),
    \label{eq:L_rad}
\end{equation}
where $\kappa$ is the opacity, $c$ the speed of light, and $a$ the radiation constant. 
When the oscillation is nearly adiabatic, the Lagrangian perturbations to the temperature $\Delta T/T$, density $\Delta \rho/\rho$, and opacity $\Delta \kappa/\kappa$ all scale with the perturbation pressure $\Delta P/P$.
Because the pressure (and hence temperature) perturbation is nearly constant near the photosphere,
\begin{align}
    \frac{\Delta L}{L_{\rm rad}} &\approx \left( 4 \frac{\Delta T}{T} - \frac{\Delta \kappa}{\kappa} - \frac{\Delta \rho}{\rho} \right) \sim \frac{\Delta P}{P}. \label{eq:dL_over_Lrad}
\end{align}
When the pressure perturbation rises, the mode causes larger luminosity fluctuations.  We note that our pressure perturbation does not vanish at the surface.  This is because we match our solutions to that of a pressure wave propagating through an isothermal atmosphere \citep[e.g.][]{Unno+(1989), Christensen-Dalsgaard2008}, using the $\texttt{outer\_bound}=\texttt{JCD}$ option in \texttt{GYRE} (App.~\ref{app:inlist}).  See Appendix~\ref{app:isotherm} for further discussion of modes above the photosphere.

\section{Mode Amplitudes}
\label{sec:ModeAmp}

An impactor impulsively imparts momentum near the surface of the planet.  Because pressure modes with a large number of radial nodes have large amplitudes near the surface (Fig.~\ref{fig:ModeProfs}), impacts preferentially excite high-frequency oscillations.  Oscillation amplitudes that are too high will result in shocks. 
The goal of this section is to estimate the oscillation amplitudes of planetary normal modes excited by the impactor's momentum, limited by 
shocks. 

\begin{figure}
\includegraphics[width=\linewidth]{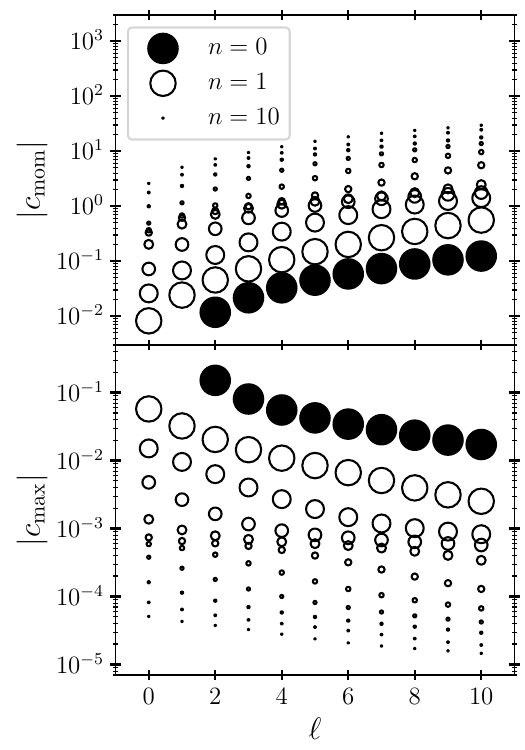}
\caption{Mode amplitudes, with $f$-modes displayed as large, filled circles, and $p$-modes as open circles, with symbol size decreasing as $n$ increases (from $n=1$ to $n=10$), vs.~angular degree $\ell$.  \textit{Top panel}: Mode amplitudes just after a merger with a Neptune-mass impactor $m_{\rm imp} = M_{\rm Nep}$ (eq.~\ref{eq:c_mom}, no damping applied).  \textit{Bottom panel}: Maximum mode amplitude set by shocks (eq.~\ref{eq:c_max}).  Notice $|c_{\rm mom}|$ and $|c_{\rm max}|$ have opposite trends with $\ell$ and $n$.  The mode amplitude after impact is given by $c = \min(c_{\rm mom}, c_{\rm max})$.
\label{fig:amps}}
\end{figure}

A giant planet's displacement from equilibrium $\bxi(r,\theta,\vphi,t)$ may be decomposed into eigenmodes $\widetilde \bxi_{\ell m n} (r,\theta,\vphi) \e^{-\im \omega_{\ell m n} t}$ projected onto vector spherical harmonics, with radial displacement $\widetilde \bxi_{r, \ell m n}(r,\theta,\vphi) = \widetilde \xi_{r,\ell m n}(r) Y_{\ell m}(\theta, \vphi) \he_r$ and horizontal displacement $\widetilde \bxi_{\perp, \ell m n} (r,\theta,\vphi) = \widetilde \xi_{\perp, \ell m n}(r) r \bdel Y_{\ell m}(\theta, \vphi)$, multiplied by a dimensionless, time-dependent coefficient 
$c_{\ell m n}(t)$, i.e.
\begin{align}
    \bxi(r,\theta,\vphi,t) = &\sum_{\ell=0}^{\infty} \sum_{m=-\ell}^{\ell} \sum_n c_{\ell mn}(t) \big[ \widetilde \xi_{r,\ell m n}(r) Y_{\ell m}(\theta, \vphi) \he_r  
    \nonumber \\
    &+ \widetilde \xi_{\perp, \ell m n}(r) r \bdel Y_{\ell m}(\theta, \vphi) \big].
    \label{eq:bxi_sum}
\end{align}
The $c_{\ell m n}$ coefficients in~\eqref{eq:bxi_sum} allow us to calculate how modes are impulsively excited after impact. Each oscillation causes Lagrangian perturbations to scalar quantities $X$ (temperature, pressure, density, etc.) proportional to 
\begin{equation}
    \Delta X(r,\theta,\vphi,t) = \sum_{\ell=0}^\infty \sum_{m=-\ell}^\ell \sum_n c_{\ell m n}(t) \Delta \widetilde X_{\ell m n}(r) Y_{\ell m}(\theta, \vphi).
\end{equation}
Note that while $\Delta X$ are proportional to a sum over $c_{\ell m n}$, the same proportionality does not apply to $\Delta \widetilde X_{\ell m n}$; all tilde-variables, including $\widetilde \bxi_{\ell m n}$, should be thought of as ``unit basis vectors''. Individual $\ell, m, n$ modes (and their complex conjugates) have energies $\eps = 2 \om^2_{\ell m n} |c_{\ell m n}|^2 \langle \widetilde \bxi_{\ell m n} | \widetilde \bxi_{\ell m n} \rangle$, where the notation $\langle {\bm u} | {\bm v} \rangle = \int {\bm u}^* \bcdot {\bm v} \rho \der V$ \citep[e.g.][]{Schenk+(2001)}.  We normalize our eigenvectors $\widetilde \bxi_{\ell m n}$ so that
\begin{align}
    &E_{\rm p} = \frac{G \Mp^2}{\Rp} = 2 \omega^2_{\ell m n} \langle \widetilde \bxi_{\ell m n} | \widetilde \bxi_{\ell m n} \rangle \nonumber \\
    &= 2 \omega^2_{\ell m n} \int_0^{R_{\rm p}} \big[ |\widetilde \xi_{r, \ell m n}|^2 + \ell(\ell+1) |\widetilde \xi_{\perp, \ell m n}|^2 \big] \rho r^2 \der r
    \label{eq:norm}
\end{align}
From here on, we drop the $\ell, m, n$ subscripts (excepting spherical harmonics), and displacement vectors will always refer to a vector spherical harmonic ($\widetilde \bxi_r = \widetilde \xi_r Y_{\ell m} \he_r$, $\widetilde \bxi_\perp = \widetilde \xi_\perp r \bdel Y_{\ell m}$).  Because the planet's rotation frequency $\Omega_{\rm p} \ll \omega_{\rm dyn} = (G \Mp/\Rp^3)^{1/2} \lesssim \omega$ for $f$-modes and $p$-modes, we will neglect Coriolis forces \citep[e.g.][]{Lai(2021), Dewberry+(2021), Dewberry+(2022), DewberryLai(2022), Lin(2023)}.

In what follows, we quantitatively calculate mode amplitudes after a giant impact.  Section~\ref{subsec:amp_mom} calculates oscillations excited by the impact's momentum $c_{\rm mom}$, while Section~\ref{subsec:amp_shock} calculates the maximum oscillation amplitude $c_{\rm max}$ before acoustic waves undergo shocks  (Sec.~\ref{subsec:amp_mom}). The value of $c$ directly after a collision is
\begin{equation}
    c = \min(c_{\rm mom}, c_{\rm max}).
    \label{eq:c_amp}
\end{equation}
Section~\ref{sec:DispTempAmp} calculates the radial displacement and temperature perturbation at the photosphere.  Except for low $\ell$ and $n$, most normal modes are strongly non-linear after impact. Heat imparted by the collision can also drive oscillations, but the dynamics are more sensitive to the uncertain thermal properties of young planets.  We provide optimistically large estimates for mode amplitudes excited by impactor heating in Appendix~\ref{app:heat}, and find them to be comparable in magnitude to those excited by the impactor's momentum \citep[see also][]{DombardBoughn(1995)}.

\subsection{Momentum Imparted After Impact}
\label{subsec:amp_mom}

When a mode is externally driven by a force per unit mass ${\bm f}$, its amplitude evolves as
\begin{equation}
    \dot c + \im \om c + \cg c  =  \frac{\im \omega}{E_{\rm p}} \langle \widetilde \bxi | {\bm f} \rangle,
    \label{eq:dot_c}
\end{equation}
where $\gamma>0$ is the mode damping rate \citep[e.g.][]{Schenk+(2001),LaiWu(2006)}.  We assume a body with mass $m_{\rm imp}$ and velocity $v_{\rm esc} = \sqrt{2 G \Mp/\Rp}$ impulsively imparts momentum in the radial direction, with a force per unit volume
\begin{equation}
    \rho f_{\rm mom} = - m_{\rm imp} v_{\rm esc} \delta^{(3)}(\br - \br_{\rm imp}) \delta(t) \he_r,
    \label{eq:f_mom}
\end{equation}
where $\delta^{(3)}({\bm r})$ is a 3-dimensional $\delta$-function, and $\delta(t)$ is a 1-dimensional $\delta$-function\citep[e.g.][]{Lognonne+(1994), DombardBoughn(1995), WuLithwick(2019)}. Taking the inner product of~\eqref{eq:f_mom} with $\widetilde \bxi$, we have
\begin{align}
    &\langle \widetilde \bxi | {\bm f}_{\rm mom} \rangle = - m_{\rm imp} v_{\rm esc} \delta(t) \widetilde \xi_r(\Rp) Y_{\ell m}(\theta_{\rm imp}, \vphi_{\rm imp}) 
\end{align}
where ($\theta_{\rm imp}$, $\vphi_{\rm imp}$) are the latitudinal and azimuthal coordinates of the impact. If we take the impact to occur at the pole ($\theta_{\rm imp}=0$; this can be assumed without loss of generality when the planet's rotation is negligible), only $m=0$ modes have non-zero amplitudes after impact of
\begin{equation}
    c_{\rm mom} = - \im \frac{m_{\rm imp} v_{\rm esc} \omega}{E_{\rm p}} \widetilde \xi_r(\Rp) \sqrt{ \frac{2\ell + 1}{4\pi} }.
    \label{eq:c_mom}
\end{equation}
The factor $\sqrt{(2\ell + 1)/(4\pi)}$ comes from evaluating $Y_{\ell 0}|_{\theta_{\rm imp}=0}$, and $\widetilde \xi_r(R_{\rm p})$ is determined after normalizing the oscillation using equation~\eqref{eq:norm}.
Equation~\eqref{eq:c_mom} implies that $c_{\rm mom}$ is linearly proportional to the mass of the impactor, $m_{\rm imp}$.
Because larger $\ell$ and $n$ modes have less inertia, $|c_{\rm mom}|$ increases with the angular degree and the number of radial nodes of a mode (Fig.~\ref{fig:amps} top panel).  

From here on, because we fix $\theta_{\rm imp} = 0$, we will only consider zonal, axisymmetric oscillations ($m=0$).

\subsection{Shock Amplitude Limit}
\label{subsec:amp_shock}
When propagating through a medium with sound-speed $c_{\rm s}$, sound waves 
develop into shocks 
when their velocities $|{\bm v}_{\rm mode}| \approx \omega |\bxi| \gtrsim c_{\rm s}$.  Because sound waves have radial wave-numbers $k_r \approx \omega/c_{\rm s}$ \citep[e.g.][]{Unno+(1989)}, the shock condition translates to 
\begin{equation}
    |{\bm k} \bcdot {\bm \xi}| \approx \frac{\partial}{\partial r}(\he_r \bcdot \bxi) \gtrsim 1.
\end{equation}
Therefore, the maximum amplitude that modes can reach before they shock is given by
\begin{equation}
    c_{\rm max} \approx \left| \frac{\der \widetilde \bxi_r}{\der r} \right|^{-1}_{\rm max} \approx \sqrt{ \frac{4\pi}{2 \ell + 1} } \left| \frac{\der \widetilde \xi_r}{\der r} \right|^{-1}_{\rm max},
    \label{eq:c_max}
\end{equation}
where $|\dots|_{\rm max}$ is the maximum amplitude over $r$, and the factor of $\sqrt{4\pi/(2\ell + 1)}$ comes from evaluating $\widetilde \bxi_r|_{\theta_{\rm imp}=0} = \widetilde \xi_r Y_{\ell m}|_{\theta_{\rm imp}=0}$.  The magnitude of $|c_{\rm max}|$ decreases with both $\ell$ and $n$ (Fig.~\ref{fig:amps} bottom panel), because these modes are more compressional and have higher $|\der \widetilde \xi_r/\der r|$.

\subsection{Displacement and Temperature Amplitudes After Impact}
\label{sec:DispTempAmp}

\begin{figure}
\includegraphics[width=\linewidth]{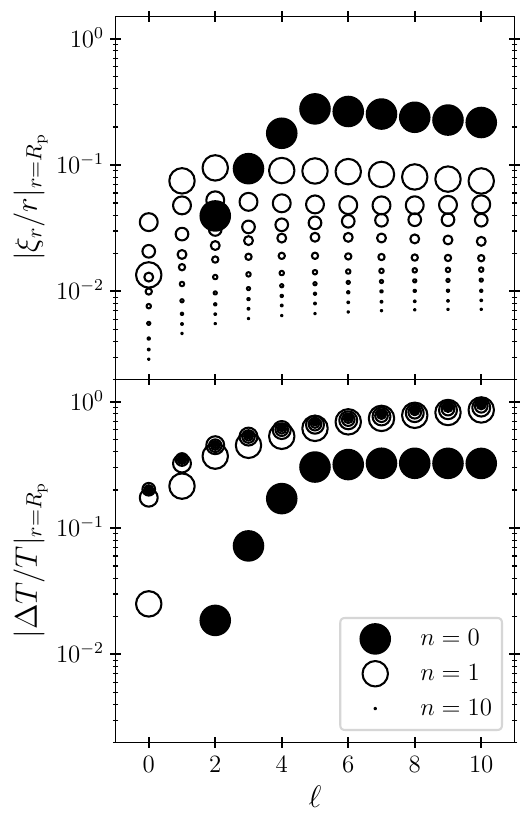}
\caption{
The radial displacement (top panel) and temperature variation (bottom panel) at the surface just after impact (no damping applied), versus angular degree.  Large, filled circles denote $f$-modes, while open circles denote $p$-modes, with symbol size decreasing as $n$ increases (from $n=1$ to $n=10$).  Modes with high $\ell,n$ are strongly non-linear ($|\partial \xi_r/\partial r| \sim 1$), with order unity variations in $|\Delta T/T|$.
\label{fig:disp&temp}
}
\end{figure}

The amplitude of the radial displacement and Lagrangian temperature perturbation for each $(\ell, m, n)$ normal mode is
\begin{align}
    \xi_r(r,\theta,\vphi,t) &= 2 c(t) \, \widetilde \xi_r(r) \, Y_{\ell m}(\theta,\vphi),
    \label{xir_phys}\\
    \Delta T(r,\theta,\vphi,t) &= 2c(t) \, \Delta \widetilde T(r) \, Y_{\ell m}(\theta,\vphi).
    \label{DT_over_T}
\end{align}
We multiply by factors of $2c$ (rather than $c$), because eigenmodes are added to their physically identical complex-conjugate.  At the impact time $t=0$, $c(0)$ is given by equation~\eqref{eq:c_amp}.  Figure~\ref{fig:disp&temp} displays the $|\xi_r/r|$ and $|\Delta T/T|$ amplitudes at the pole on the planet's surface $(\{r, \theta\} = \{R_{\rm p}, 0\})$.  Only $n=0$ $f$-modes with $\ell < 4$, and $n=1$ $p$-modes with $\ell < 2$, have amplitudes set by momentum imparted by the impactor ($c = c_{\rm mom}$).  Higher $\ell, n$ oscillations are shock-limited, with temperature variations clustering around $|\Delta T/T| \sim |\Delta \rho/\rho| \sim |k_r \xi_r| \sim 1$.  These order unity $|\Delta T/T|$ amplitudes call into question the validity of our linear treatment.  However, the longer one waits, the more low $\ell, n$ normal modes dominate the oscillation spectrum, for which our linear treatment is applicable.  In the next section, we will show that high $\ell, n$ modes have short lifetimes, and only low $\ell, n$ oscillations persist over a timescale comparable to the planet's age.

\section{Internal Dissipation}
\label{sec:ModeDamp}

\begin{figure}
\includegraphics[width=\linewidth]{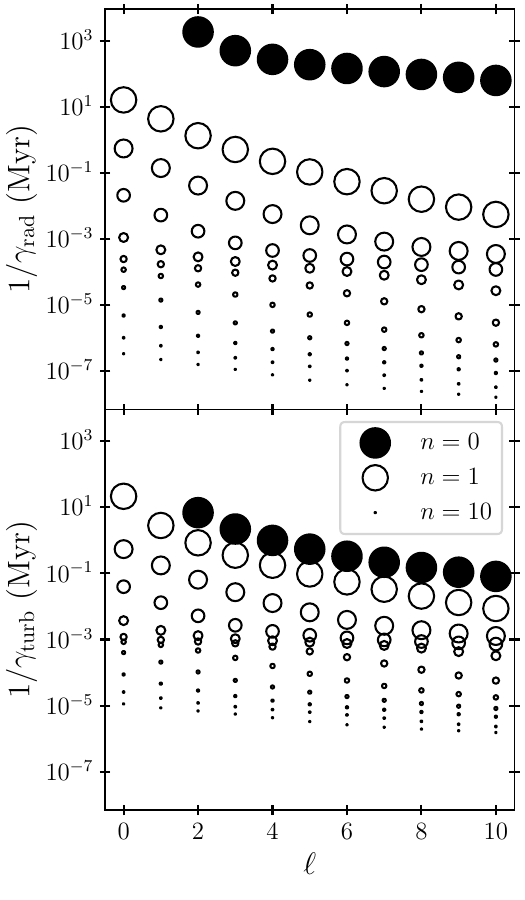}
\caption{Mode damping timescales $1/\gamma$, with $f$-modes displayed as large, filled circles, and $p$-modes as open circles, with symbol size decreasing as $n$ increases (from $n=1$ to $n=10$), versus angular degree $\ell$. 
 \textit{Top panel}: Damping from radiative diffusion.  \textit{Bottom panel}: Damping from turbulent convection.  Notice that the two damping timescales tend to be comparable in magnitude.
\label{fig:damp}}
\end{figure}

Once excited, normal modes damp from internal friction within the giant planet.  Here, we calculate the linear damping rates $\gamma$ for $f$ and $p$-modes from radiative losses and turbulent convection.  In agreement with \cite{WuLithwick(2019)}, we find both processes to be sufficiently weak that the longest period $p$-mode and $f$-modes decay over the global Kelvin-Helmholtz timescale $\gamma^{-1} \sim E_{\rm p}/L \sim 10 \ {\rm Myr}$ (Fig.~\ref{fig:damp}). We note, however, that our linear damping rates should be taken as lower bounds, because the steep profiles for many normal modes are such that 
$|\partial \bxi_r/\partial r| \sim 10^{-2} - 1$ after the merger, implying non-linear effects could be important (Fig.~\ref{fig:amps}; see e.g. \citealt{KumarGoldreich(1989), Kumar+(1994), KumarGoodman(1996), MacLeod+(2022)}). 


\subsection{Radiative Damping}
\label{sec:DampRad}

As the planet oscillates, radiation slowly saps $f$ and $p$-modes of their energy.  The mode perturbs the radiative flux ${\bm F}_{\rm rad}$, causing the Lagrangian perturbation to the entropy $\Delta S$ to be non-zero:
\begin{equation}
    T \frac{\partial}{\partial t} \Delta S = - \Delta \left( \frac{1}{\rho} \bdel \bcdot {\bm F}_{\rm rad} \right).
    \label{eq:entropy_change}
\end{equation}
When the thermal time $t_{\rm th}$ is much longer than $\om^{-1}$, $\Delta S$ can be calculated perturbatively, in the so-called  `quasi-adiabatic approximation' \citep{Unno+(1989)}.  To leading order, density, temperature, and opacity variations are adiabatic, and proportional to the pressure perturbation:
\begin{equation}
    \frac{\Delta T}{T}, \frac{\Delta \rho}{\rho}, \frac{\Delta \kappa}{\kappa} \propto \frac{\Delta P}{P}.
\end{equation}
Adiabatic variations in $\Delta P/P$ can then be used to calculate the mode's luminosity variation, by taking the Lagrangian perturbation $\Delta L$ of equation~\eqref{eq:L_rad}. The variation in luminosity is then inserted into equation~\eqref{eq:entropy_change}, to find $\Delta S$ to leading order.

The mode energy decreases from 
radiative diffusion over the volume of the planet, 
and from work done by the oscillation at the planet's photosphere (above which radiation freely streams):
\begin{equation}
    \dot \eps_{\rm rad} = \dot \eps_{\rm diff} + \dot \eps_{\rm photo},
\end{equation}
where
\begin{align}
    &\dot \eps_{\rm diff} = \int \int_{r=0}^{\Rp} \left( \Dg T^* \frac{\pd}{\pd t} \Dg S + \Dg T \frac{\pd}{\pd t} \Dg S^* \right)  \rho r^2 \der r \, \der \Om
    \nonumber \\
    &= 2 \Re \left( \int \int_{r=0}^{\Rp} \Delta T^* \frac{\pd}{\pd t} \Delta S \, \rho r^2 \der r \, \der \Omega \right), 
    \label{eq:dotE_diff} \\
    &\dot \eps_{\rm photo} = -\left[ \int \left( \Delta P^* \frac{\pd}{\pd t} \bxi + \Dg P \frac{\pd}{\pd t} \bxi^*\right)\bcdot \he_r \, r^2 \der \Omega \right]_{r=\Rp} \nonumber \\
    &= - 2 \Re \left[ \int \Delta P^* \left(\frac{\pd}{\pd t} \bxi\right) \bcdot \he_r \, r^2 \der \Omega \right]_{r=\Rp}
\end{align}
We set the photospheric radius $R_{\rm p}$ to be where $\tau = 2/3$. In $\dot \eps_{\rm rad}$, each mode is added to its complex conjugate.  
Because the damping rate $|c(t)| \propto \e^{-\gamma t}$ is related to the change in mode energy by $\dot \eps = 2 \gamma \eps$, the damping rates due to diffusion in the interior and ``leakage'' through the photosphere are, respectively,
\begin{align}
    &\gamma_{\rm diff} = \frac{\dot \eps_{\rm diff}}{2\eps} = \frac{\om}{E_{\rm p}} \int_0^{\Rp} \Im \left( \Dg \widetilde T^* \Dg \widetilde S \right) \rho r^2 \der r, 
    \label{eq:gamma_diff} \\
    &\gamma_{\rm photo} = \frac{\dot \eps_{\rm photo}}{2\eps} = - \frac{\om}{E_{\rm p}} \left[ \Im \left( \Dg \widetilde P^* \widetilde \xi_r  \right) r^2  \right]_{r=\Rp}.
    \label{eq:gamma_surf}
\end{align}
To derive~\eqref{eq:gamma_diff} and~\eqref{eq:gamma_surf}, we integrate over solid angle, and quantities $X Y$ become $\widetilde X \widetilde Y$ after dividing by $\eps \propto |c|^2$.
To calculate $\cg_{\rm diff}$, we use the \texttt{GYRE} code to calculate $\Delta \widetilde S$ and $\Delta \widetilde T$ under the quasi-adiabatic approximation as described above, and then integrate equation~\eqref{eq:gamma_diff} numerically. 

We show in Appendix~\ref{app:isotherm} that for modes in a plane-parallel, isothermal atmosphere, $\cg_{\rm photo}$ is given by
\begin{equation}
    \gamma_{\rm photo} \simeq \left( \frac{4 \upsilon_T \nabla_{\rm ad} P r^2 F_{\rm rad}}{\rho C_P T E_{\rm p}} \left| \frac{\Dg \widetilde P}{P} \right|^2 \right)_{r=\Rp},
    \label{eq:gamma_photo}
\end{equation}
where $\upsilon_T = -(\pd \ln \rho/\pd \ln T)_S$ and $\nabla_{\rm ad} = (\pd \ln T/\pd \ln P)_S$.
The total damping rate due to radiation $\gamma_{\rm rad} = \gamma_{\rm diff} + \gamma_{\rm photo}$ is plotted in the top panel of Figure~\ref{fig:damp}. It is smallest for $f$-modes and low-$n$ $p$-modes.

The contributions to $\cg_{\rm diff}$ are greatest near the photosphere where energy is transported predominantly by radiative diffusion: we neglect the entropy change of the oscillation from convection (``Case 1'' of \citealt{Pesnell1990}).  
We can obtain an order-of-magnitude estimate of $\gamma_{\rm diff}$ as follows.  The total internal luminosity is dominated by radiation 
one pressure scale-height $H_P$ 
below the surface (i.e.~$L \simeq L_{\rm rad}$ when $r \gtrsim \Rp - H_P$), where pressure, temperature, and density drop sharply ($H_P/r \ll 1$). Here, the 
right-hand side of \eqref{eq:entropy_change} can be estimated as 
\begin{equation}
  T \frac{\pd}{\pd t} \Delta S \sim \frac{\Delta \rho}{\rho} \frac{F_{\rm rad}}{\rho H_P}.
\end{equation}
Because adiabatic pulsations satisfy
\begin{equation}
    \frac{\Delta T}{T} \sim \frac{\Delta \rho}{\rho} \sim \frac{\Delta P}{P},
\end{equation}
and the radiative flux at the surface is of order $F_{\rm rad} \sim L/r^2$, the diffusive damping rate is of order 
\begin{align}
    \gamma_{\rm diff} &\sim \frac{1}{\eps} \int\int_{\Rp - H_P}^{\Rp} \frac{F_{\rm rad}}{\rho H_P} \left| \frac{\Delta T}{T} \right|^2 \rho \, r^2 \der r \, \der \Omega
    \nonumber \\
    &\sim \int_{\Rp-H_P}^{\Rp}  \frac{L}{E_{\rm p}} \left| \frac{\Delta \widetilde T}{T} \right|^2 \frac{\der r}{H_P}.
    \label{eq:gamma_rad_mag}
\end{align}
Low-order $p$-modes and $f$-modes 
have $|\Delta \widetilde T/T| \sim 1$.  
Since the integrand is roughly constant near the photosphere ($\der r \sim H_P$),
the radiative damping rate should be of order the inverse Kelvin-Helmholtz timescale $L/E_{\rm p}$. 
Because higher frequency modes are more 
compressional ($|\Delta \widetilde T/T|$ increases), $\gamma_{\rm diff}$ increases with $\om$.

The order of magnitude and scaling of $\gamma_{\rm photo}$ are similar to those of $\gamma_{\rm diff}$.   Because the radial gradient of the temperature perturbation near the planet's photosphere goes like $({\pd}/{\pd r} )(\Delta T/T) \sim ({\Lambda_-}/{H_P})(\Delta T/T)$, where $\Lambda_- \sim \om^2/\om_{\rm ac}^2$ is given by equation~\eqref{eq:Lambda_-}, and $\om_{\rm ac} = c_{\rm s}/(2 H_P) \sim 2\pi/(3 \, {\rm min})$ is the acoustic cutoff frequency, the heat lost per time per mass is of order 
\begin{equation}
    T \frac{\pd}{\pd t} \Dg S \sim -\frac{\Lambda_- F_{\rm rad}}{\rho H_P} \frac{\Dg T}{T}.
    \label{eq:entropy_photo_mag}
\end{equation}
The change in density can be related to the change in entropy and temperature using the laws of thermodynamics:
\begin{equation}
    \frac{\Dg \rho}{\rho} \sim -\frac{\Lambda_-}{H_P} \xi_r \sim \frac{\Dg T}{T} - \frac{\Dg S}{C_P}.
    \label{eq:xir_thermo_mag}
\end{equation}
Inserting equations~\eqref{eq:entropy_photo_mag} and~\eqref{eq:xir_thermo_mag} into~\eqref{eq:gamma_surf} then gives
\begin{align}
    &\gamma_{\rm photo} \sim -\frac{1}{\eps} \left[ \int P \frac{\Dg T^*}{T} \frac{H_P}{\Lambda_-} \frac{\pd}{\pd t}  \left( \frac{\Dg S}{C_P} \right) r^2 \der \Om \right]_{r=\Rp}
    \nonumber \\
    &\sim \frac{1}{E_{\rm p}} \left( \frac{P F_{\rm rad} r^2}{\rho C_P T} \left| \frac{\Dg \widetilde T}{T} \right|^2 \right)_{r=\Rp} \sim \frac{L}{E_{\rm p}} \left| \frac{\Dg \widetilde T}{T} \right|^2_{r = \Rp},
\end{align}
where we have used $P \sim \rho C_P T$.  For low angular degree $f$-modes and $n=1$ $p$-modes,  $|\Dg \widetilde T/T|_{r=\Rp} \sim 1$, and $\cg_{\rm photo}$ is of order the inverse Kelvin-Helmholtz timescale $L/E_{\rm p}$. Modes with higher frequencies become more compressional ($|\Dg \widetilde T/T|_{r = \Rp}$ increases), which increases $\cg_{\rm photo}$.

\subsection{Turbulent Damping}
\label{sec:DampTurb}

In convective regions, turbulent eddies sap the mode energy at a rate proportional to the mode's shear, much like an effective viscosity.  This causes the mode to lose energy at the rate
\begin{align}
    \cg_{\rm turb} &= \frac{\dot \eps_{\rm turb}}{2\eps} = \frac{\omega^2}{ \eps} \int  \int_0^{\Rp}\nu_{\rm turb} \bdel \bxi {\bm :} \bdel \bxi \rho r^2 \der r \, \der \Omega \nonumber \\
    &= \frac{\om^2}{E_{\rm p}} \int_0^{\Rp} \rho \nu_{\rm turb} X_\ell^2(r) \der r,
    \label{eq:gamma_turb}
\end{align}
where \citep[e.g.][]{Barker(2020)}
\begin{align}
    X_\ell^2 &= 3 \left| r \frac{\der \widetilde \xi_r}{\der r} - \frac{r \Delta_\ell}{3} \right|^2 + \ell(\ell+1) \left| \widetilde \xi_r + r \frac{\der \widetilde \xi_\perp}{\der r} - \widetilde \xi_\perp \right|^2 \nonumber \\
    &+ (\ell-1)\ell(\ell+1)(\ell+2) |\widetilde \xi_\perp|^2, \\
    r \Delta_\ell &= r \frac{\der \widetilde \xi_r}{\der r} + 2 \widetilde \xi_r - \ell(\ell+1) \widetilde \xi_\perp.
\end{align}
Recent hydrodynamical simulations show one can parameterize the effective viscosity $\nu_{\rm turb}$ using mixing length theory:
\begin{equation}
    \nu_{\rm turb} = u_{\rm turb} l_{\rm turb}
    \left\{ \begin{array}{ll}
    5 & \dfrac{\om}{\om_{\rm turb}} < 0.01, \\
    \frac{1}{2} \left( \dfrac{\om_{\rm turb}}{\om} \right)^{1/2} & 0.01 \le \dfrac{\om}{\om_{\rm turb}} \le 5 \\
    \dfrac{25}{\sqrt{20}} \left( \dfrac{\om_{\rm turb}}{\om} \right)^2 & \dfrac{\om}{\om_{\rm turb}} > 5
    \end{array} \right. ,
\end{equation}
where the convective velocity $u_{\rm turb}$ and mixing length $l_{\rm turb}$ are \texttt{MESA} outputs, 
and $\omega_{\rm turb} = u_{\rm turb}/l_{\rm turb}$ is the turnover frequency \citep[e.g.][]{VidalBarker(2020a), VidalBarker(2020b), Duguid+(2020a), Duguid+(2020b)}.  Plugging our $f$ and $p$-mode \texttt{GYRE} oscillations into equation~\eqref{eq:gamma_turb}, we find $1/\gamma_{\rm turb}$ is longest for $f$-modes and low $n$ $p$-modes, sometimes exceeding several million years (Fig.~\ref{fig:damp}).

The dissipation from turbulent damping is largest near the planet's surface.  Because for our high-frequency modes, $\omega \gg \omega_{\rm turb}$ everywhere,
\begin{equation}
    \nu_{\rm turb} \sim \frac{F_{\rm conv}}{\omega^2 \rho H_P},
    \label{eq:nu_turb_est}
\end{equation}
where the flux $F_{\rm conv} \sim \rho u_{\rm turb}^3 \sim L/r^2$ in convective regions, with mixing lengths of order $l_{\rm turb} \sim H_P$.  We can use \eqref{eq:nu_turb_est} to understand the magnitude of the turbulent damping rate, by inserting it into equation~\eqref{eq:gamma_turb}, and using $X_\ell^2 \sim | \der \widetilde \xi_r/\der \ln r|^2$:
\begin{align}
    \gamma_{\rm turb} &\sim \frac{1}{\eps} \int \int_0^{\Rp - H_P} \frac{F_{\rm conv}}{\rho H_P} \left|\frac{\partial \xi_r}{\partial r} \right|^2 \rho r^2 \der r \, \der \Omega
    \nonumber \\
    &\sim \int_0^{\Rp - H_P} \frac{L}{E_{\rm p}} \left| \frac{\der \widetilde \xi_r}{\der r} \right|^2 \frac{\der r}{H_P}.
    \label{eq:gamma_turn_est}
\end{align}
Similar to mode damping from radiative diffusion, turbulent damping causes low-order $f$ and $p$-modes to decay at rates of order the inverse Kelvin-Helmholtz timescale $L/E_{\rm p}$.  As the mode frequency increases, so does its shear $|\der \widetilde \xi_r/\der r|$ (Fig.~\ref{fig:ModeProfs}), shortening the damping timescale $\gamma_{\rm turb}^{-1}$ (Fig.~\ref{fig:damp} bottom panel).

\section{Luminosity Variations after Impact}
\label{sec:LumVar}

\begin{figure*}
\centering
\includegraphics[width=0.8\linewidth]{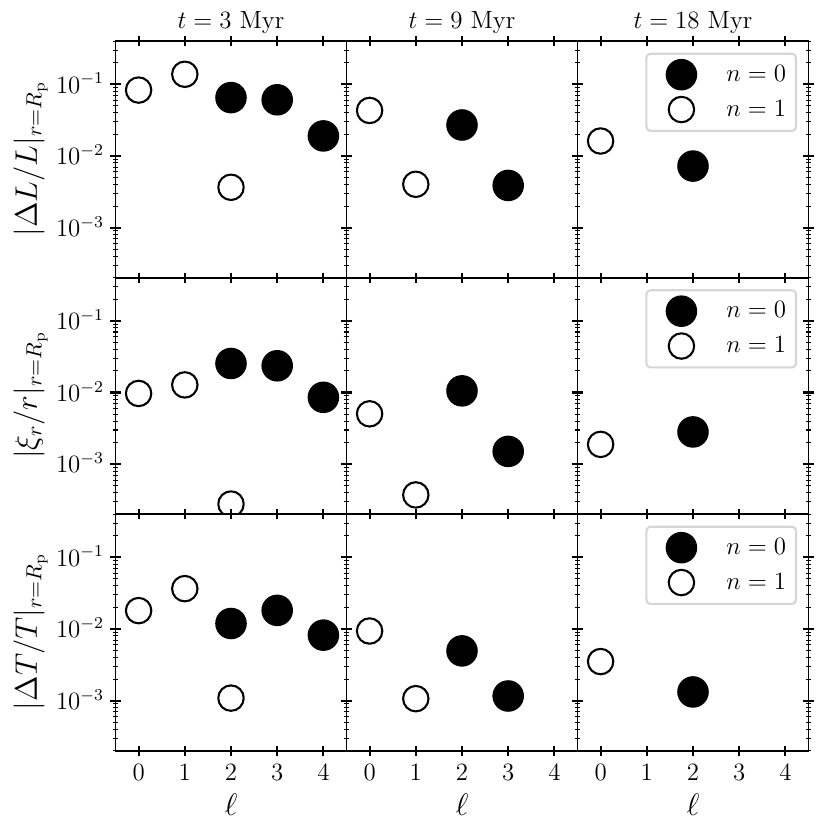}
\caption{Fractional bolometric luminosity variation (top panels), radial displacement (middle panels), and fractional temperature perturbation (bottom panel) at the surface after a merger with a Neptune-mass body, for modes with radial node nmbers $n=0$ (larger closed circles) and $n=1$ (smaller open circles), varying angular degree $\ell$.  Modes with $n>1$ vary surface $|\Delta L/L|$, $|\xi_r/r|$, and $|\Delta T/T|$ at amplitudes smaller than $10^{-4}$.  Different columns display different lengths of time after the merger. After impact, numerous normal modes persist over timescales comparable to the age of $\beta$ Pic b.}
\label{fig:osc_versus_t}
\end{figure*}

If internal friction is sufficiently weak, $f$-modes and $p$-modes excited by a merger persist over a significant fraction of a young planet's life, causing the luminosity to vary over the mode periods.  In this section, we estimate the magnitude of these luminosity variations for our fiducial case of a merger between $\beta$ Pic b and a Neptune-mass impactor. Mode amplitudes are initially $|c(0)| = \min(|c_{\rm mom}|, |c_{\rm max}|)$ (eq.~\ref{eq:c_amp}), and evolve with time according to $|c(t)| \propto \exp(-\gamma t)$, where the damping rate $\gamma = \gamma_{\rm rad} + \gamma_{\rm turb}$ (Sec.~\ref{sec:ModeDamp}).  Figure~\ref{fig:osc_versus_t} calculates different mode properties at the planet's surface with time.  After a few million years, despite the initially high amplitudes of large $\ell$ and $n$ modes, their luminosity fluctuations diminish below $|\Delta L/L| < 10^{-4}$ due to their short lifetimes.  Only low $\ell, n$ modes persist over timescales comparable to the age of $\beta$ Pic b.

In what follows, we investigate in more detail the feasibility of observing pulsations using JWST photometry.  We first calculate the fluctuation amplitudes for $\beta$ Pic b normal modes visible by the JWST NIRCam instrument (Sec.~\ref{sec:IR_Mag}), followed by a simulation of a NIRCam coronagraphic observation to see if we can recover the signal of the longest-lived normal mode in $\beta$ Pic b (Sec~\ref{sec:JWST_obs}).

\subsection{Infrared Magnitude with Time}
\label{sec:IR_Mag}

\begin{figure}
\includegraphics[width=\linewidth]{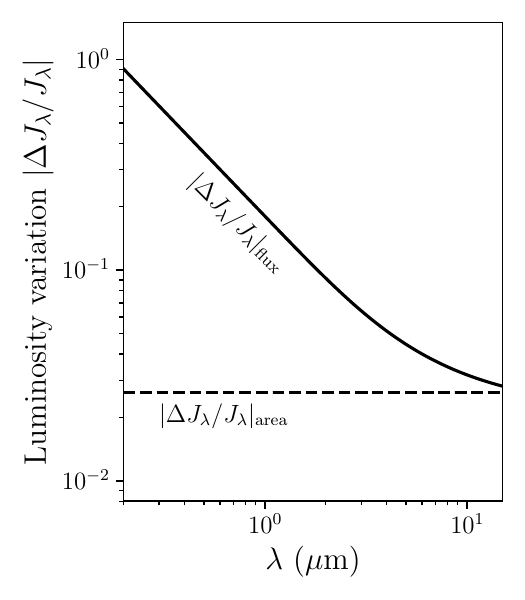}
\caption{Observed luminosity variation per unit wavelength $\Delta J_\lambda/J_\lambda$ of $\beta$-Pic b, for the $\ell=0$, $n=1$ $p$-mode after 18 Myr following a merger with a Neptune-mass body (Fig.~\ref{fig:amps}).  The luminosity variation caused by the heating and cooling of the photosphere $(\Delta J_\lambda/J_\lambda)_{\rm flux}$ (eq.~\ref{eq:lagL_flux}) dominates that from the variation of the planet's projected area onto the sky plane $(\Delta J_\lambda/J_\lambda)_{\rm area}$ (eq.~\ref{eq:lagL_area}).
\label{fig:Lum_wavelength}
}
\end{figure}

As shown in \cite{Dziembowski(1977)}, \cite{Robinson+(1982)}, \cite{Pfahl+(2008)}, and \cite{Burkart+(2012)}, a body with normal mode oscillations will have its luminosity fluctuate from two effects. First, the radial displacement causes the surface area of the planet seen by a distant observer to vary, causing the planet's observed luminosity $J$ to vary by $\Delta J$:
\begin{equation}
    \left( \frac{\Delta J}{J} \right)_{\rm area} = 2c \left( \frac{\widetilde \xi_r}{r} \right)_{\Rp} (2 b_\ell - d_\ell)  Y_{\ell m}(\theta_{\rm obs}, \vphi_{\rm obs})
    \label{eq:lagL_area_bol}
\end{equation}
where $(\theta_{\rm obs}, \vphi_{\rm obs})$ denote the angular coordinates of the observer's viewing direction,
\begin{align}
    b_\ell &= \int_0^1 \mu Q_\ell(\mu) h(\mu) \der \mu, \\
    d_\ell &= \int_0^1 \mu \left[ 2 \mu \frac{\der Q_\ell}{\der \mu} - (1-\mu^2) \frac{\der^2 Q_\ell}{\der \mu^2} \right] h(\mu) \der \mu \nonumber \\
    &= \ell(\ell+1) b_\ell
\end{align}
with $h(\mu)$ is the limb-darkening law which we take from Eddington ($h = 1 + 3\mu/2$), and $Q_\ell(\mu)$ are Legendre polynomials.  Second, the oscillation heats and cools the photosphere, which perturbs the flux:
\begin{align}
    \left( \frac{\Delta J}{J} \right)_{\rm flux} = 2c \left( \frac{\Delta \widetilde F}{F} \right)_{\Rp} b_\ell  Y_{\ell m} (\theta_{\rm obs}, \vphi_{\rm obs}),
    \label{eq:lagL_flux_bol}
\end{align}
where the bolometric flux $F = \sigma T_{\rm eff}^4$ and its Lagrangian perturbation 
$\Delta F = c \  \Delta \widetilde F \ Y_{\ell m}$ can be related to that of the luminosity using
\begin{equation}
    \frac{\Delta \widetilde F}{F} = \frac{\Delta \widetilde L}{L} - 2 \frac{\widetilde \xi_r}{r},
\end{equation}
with $\sigma$ the Stefan-Boltzmann constant. Equations~\eqref{eq:lagL_area_bol} and~\eqref{eq:lagL_flux_bol} include factors of $2c$ (rather than $c$) after adding the oscillation to its physically identical complex conjugate.  

To re-write equations~\eqref{eq:lagL_area_bol} and~\eqref{eq:lagL_flux_bol} in terms of the luminosity variation per unit wavelength $\Delta J_\lambda$, we approximate the emission as that of a blackbody with effective temperature $T_{\rm eff}$ ($J_\lambda \approx B_\lambda J/B$, with $B_\lambda$ the Planck function and $B = \sigma T_{\rm eff}^4/\pi$ the integrated Planck function; \citealt{RybickiLightman(1986)}).  Using $\Delta F_\lambda/F \approx (\partial \ln B_\lambda/\partial \ln T_{\rm eff}) \Delta T_{\rm eff}/T_{\rm eff} = (\partial \ln B_\lambda/\partial \ln T_{\rm eff}) \Delta F/(4 F)$, we have
\begin{align}
    \left(\frac{\Delta J_\lambda}{J_\lambda} \right)_{\rm area} &\approx \left( \frac{\Delta J}{J} \right)_{\rm area},
    \label{eq:lagL_area} \\
    \left( \frac{\Delta J_\lambda}{J_\lambda} \right)_{\rm flux} &\approx \frac{1}{4 B_\lambda} \frac{\partial B_\lambda}{\partial \ln T_{\rm eff}} \left( \frac{\Delta J}{J} \right)_{\rm flux} \,.
    \label{eq:lagL_flux}
\end{align}
The heating and cooling of the photosphere dominates $\Delta J_\lambda/J_\lambda$ over the changing surface area, for the relevant range of $\lambda$ (Fig.~\ref{fig:Lum_wavelength}).

\begin{figure}
\includegraphics[width=\linewidth]{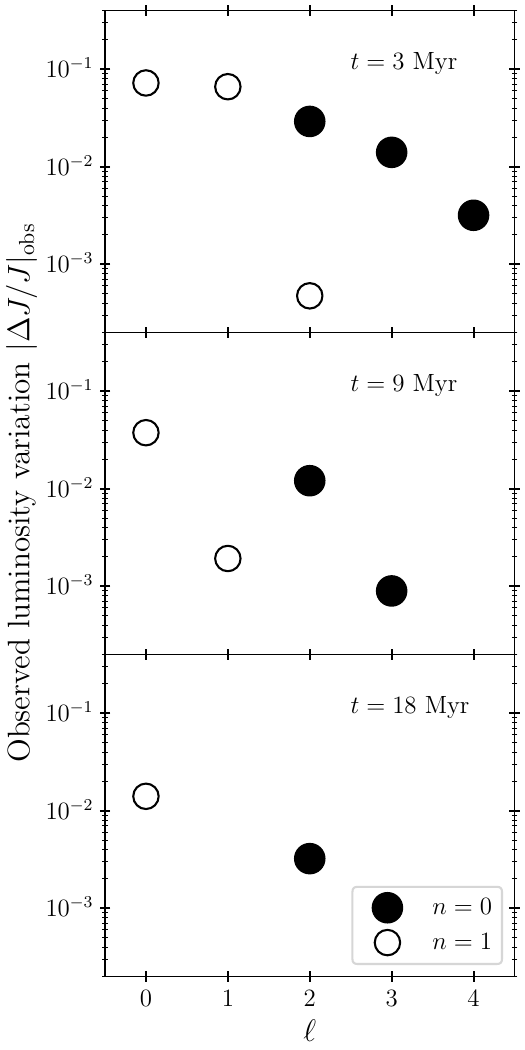}
\caption{Observed luminosity variation from $\beta$ Pic b oscillations after a merger with a Neptune-mass body, for modes with radial node numbers $n=0$ (larger closed circles) $n=1$ (smaller open circles), varying angular degree $\ell$.  Modes with $n>1$ vary the planet's luminosity with amplitudes smaller than $10^{-4}$.  Different panels display different lengths of time after the merger.  Planetary normal modes cause brightness variations whose amplitudes can exceed a percent for several million years.
\label{fig:lagL_over_L_obs}}
\end{figure}

We want to calculate the lifetime of luminosity oscillations after $\beta$ Pic b undergoes a giant impact, and the detectability of oscillations with time-series observations using JWST.  We take the JWST NIRCam instrument to be sensitive to emission from $\lambda_1 = 1.8 \ \mu{\rm m}$ to $\lambda_2 = 4.8 \ \mu{\rm m}$, so that
\begin{equation}
    \left( \frac{\Delta J}{J} \right)_{\rm obs} = \left( \frac{\Delta J}{J} \right)_{\rm area} + \beta \left( \frac{\Delta J}{J} \right)_{\rm flux},
\end{equation}
where
\begin{equation}
    \beta(T_{\rm eff}) \approx \frac{\int_{\lambda_1}^{\lambda_2} (\partial B_\lambda/\partial \ln T_{\rm eff}) \der \lambda}{4 \int_{\lambda_1}^{\lambda_2} B_\lambda \der \lambda} = 0.83
\end{equation}
accounts for how much photospheric heating contributes to $\beta$ Pic b's observed $\Delta J/J$ \citep{Burkart+(2012)}.  
Figure~\ref{fig:lagL_over_L_obs} calculates the planet's observed luminosity variations with time, after $f$ and $p$-mode oscillations are excited by a 
merger with a Neptune-mass body (Fig.~\ref{fig:osc_versus_t}). Early-on ($\lesssim$3 Myr post impact), multiple oscillations cause luminosity variations which exceed $\gtrsim$1\%.  Given sufficient time, however, one oscillation dominates the spectrum. We see the longest-lived, highest-amplitude oscillation is the $(\ell, n) = (0,1)$ $p$-mode, which causes brightness fluctuations which exceed $\gtrsim$1\% for up to $\sim$18 Myr after the impact.  In the next subsection, we investigate this mode's detectability with JWST's NIRCam instrument.   

\subsection{Detectability with JWST}
\label{sec:JWST_obs}

\begin{figure*}
\centering
\includegraphics[width=0.9\linewidth]{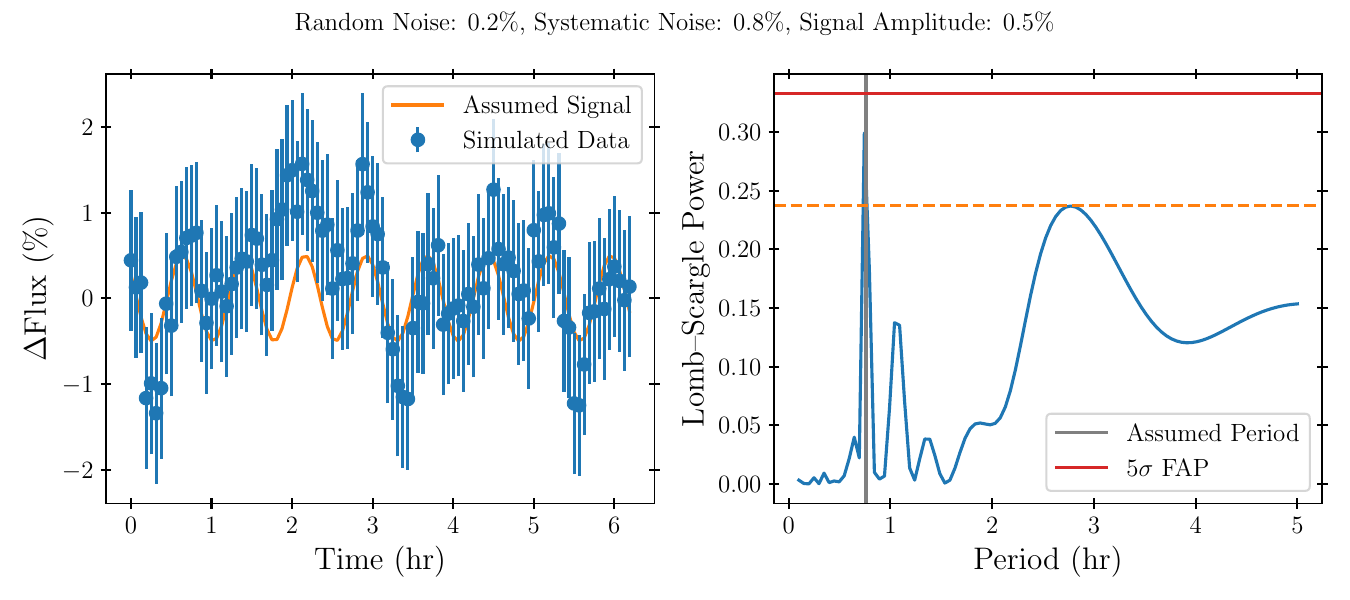}
\includegraphics[width=0.9\linewidth]{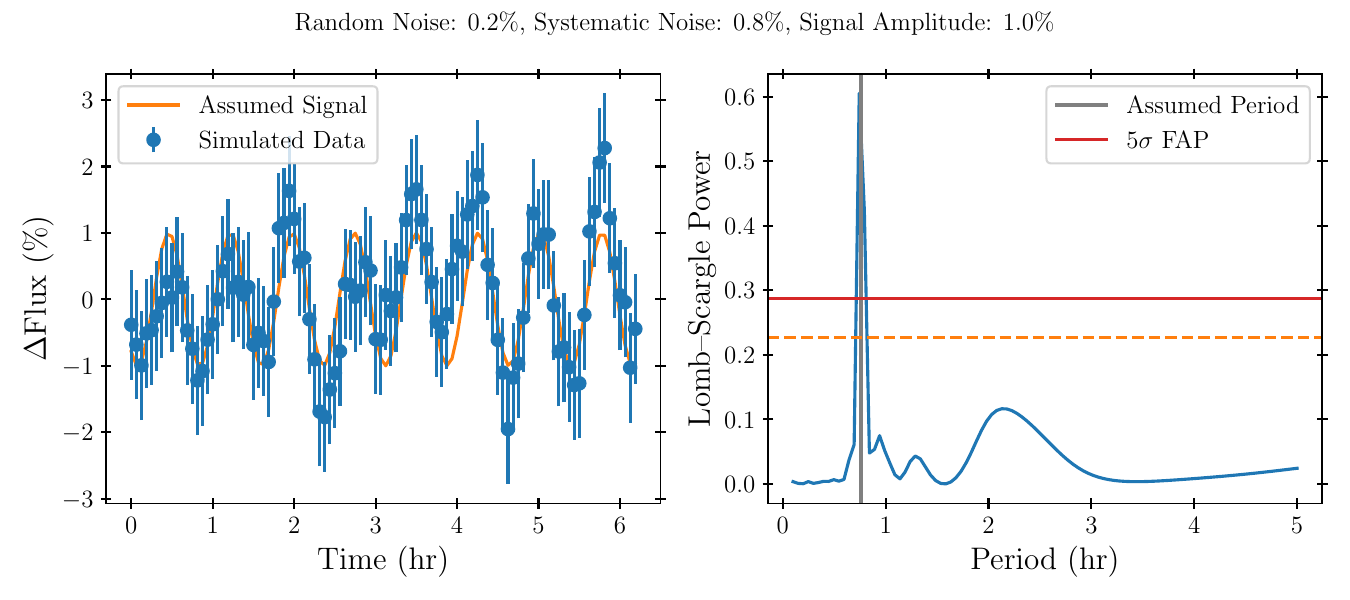}
\includegraphics[width=0.9\linewidth]{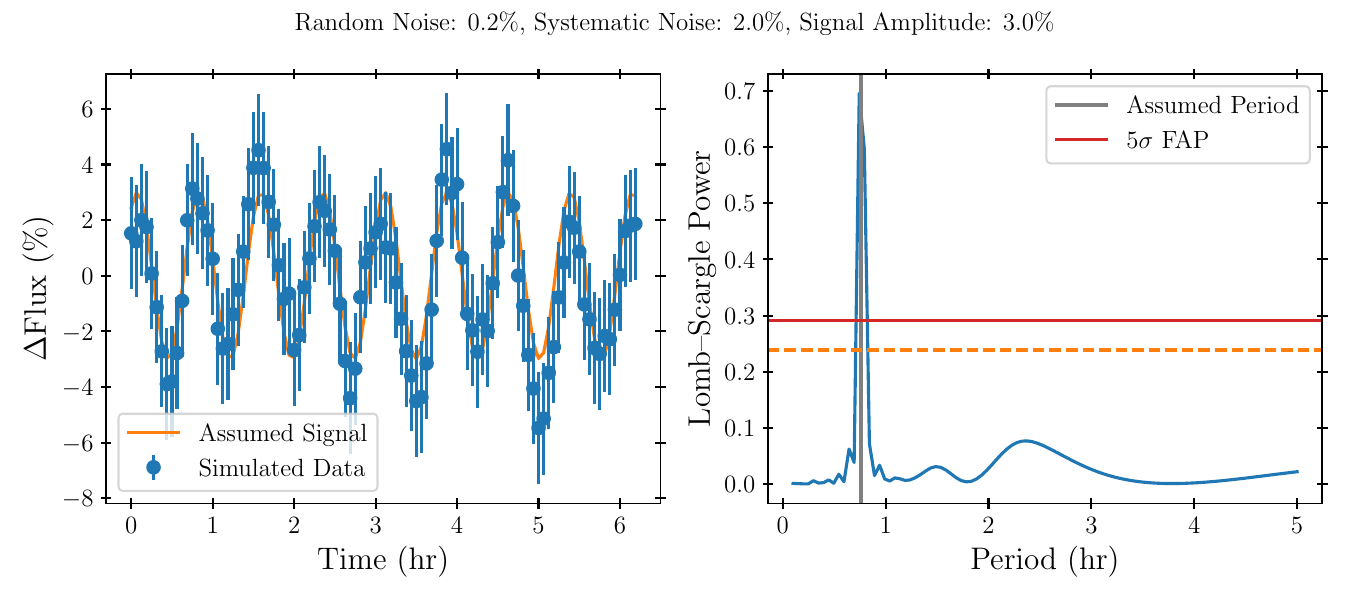}
\caption{Injecting a $\ell=0$, $n=1$ $p$-mode oscillation with period $P_{\rm mode} = 0.76 \ {\rm hr}$ into synthetic JWST data for $\beta$ Pic b, with amplitudes, random noise, and systematic noise levels indicated. \textit{Left columns}: injected signals (orange solid lines) and simulated measurements (blue dots). \textit{Right columns}: Lomb-Scargle periodograms with 3$\sigma$ (dashed lines) and 5$\sigma$ (solid lines) False Alarm Probability (FAP) upper limits indicated. 
\label{fig:betaPic_osc_JWST}}
\end{figure*}

As a demonstration, we simulate NIRCam coronagraphic observations in the F210M band to estimate the sensitivity in measuring variability for Beta Pic b. 
We inject the signal from the longest-lived $p$-mode ($\ell=0$, $n=1$, $P_{\rm mode} = 0.76$ hours, see Figs.~\ref{fig:ModeP} \&~\ref{fig:damp}), which causes $\beta$ Pic b's flux to vary with amplitudes exceeding $\sim$1\% for $\sim$18 Myr, and $\sim$3\% for $\sim$9 Myr (Fig.~\ref{fig:betaPic_osc_JWST}).
The assumed observation sequence consists of 100 consecutive exposures, each with ten BRIGHT2 groups and ten integrations. The resulting cadence and time baseline are 220 seconds and six hours, respectively. We use PanCAKE to simulate the images, perform PSF subtraction, and estimate noise \citep{Carter2021}. The validity of PanCAKE has been demonstrated in JWST direct-imaging early-release science studies \citep{Carter2023}. The F210M flux of the planet is scaled to match the F210M flux density ($f = 6.6$\,mJy) measured by \citet{Kammerer2024}. The PanCAKE simulation yields a total noise of 0.055 mJy (0.8\%), which includes photon noise of 0.015 mJy (0.2\%) and speckle noise of 0.053 mJy (0.8\%). These estimates are consistent with the output of the JWST exposure time calculator and the 5$\sigma$ sensitivity reported by \citet{Kammerer2024}.

Potential interference from stellar pulsation has been considered. Beta Pic is a $\delta$-Scuti star that is variable at the 0.1\% level \citep[e.g.,][]{Kenworthy2021}. Several oscillation modes have periods similar to those expected for the planet’s oscillation period. Stellar oscillation can impact the variability in two ways. First, it introduces modulations in the stellar PSF speckle; second, it causes the scattered light of the debris disk to vary. For the first component, assuming a pessimistic scenario where the PSF subtraction cannot de-correlate the variability of the speckle, the noise can be estimated by multiplying the stellar variability by the brightness of the PSF speckle (${\sim}$7.3\,mJy for $\beta$ Pic b), which yields 0.007 mJy (0.1\%) noise. 
Thus, if stellar pulsation amplitudes are ten times their typical level, the noise due to stellar pulsation is still less than 1\%. For the second component, although the disk brightness at the planet’s position is noticeable (100 mJy per arcsec$^2$; \citealt{Rebollido2024}), the contamination to the variability measurement is negligible because the stellar oscillation amplitude is low. In summary, we estimate the photometric random noise to be 0.2\%, the optimistic systematic noise to be 0.8\% (only speckle noise), and the pessimistic systematic noise to be 2\% (speckle and stellar variability).

We simulate the planet oscillation using a sinusoid, add noise, and try to recover it using a Lomb-Scargle periodogram. The photon noise is modeled as random noise, and the systematic noise is modeled as red noise. The significance of the detection is evaluated based on the false alarm probability estimation of the periodic signal. The recovery of the periodic signal is considered significant when the significance exceeds a false alarm probability of $2.86\times10^{-7}$ ($5\sigma$). In the optimistic scenario (0.8\% red noise), a 1\% planet oscillation can be confidently recovered; in the pessimistic scenario (2\% red noise), a 3\% planet oscillation can be recovered.

\section{Summary and Discussion}
\label{sec:SummDisc}

The vast stores of heavy metals in Jupiter-mass exoplanets \citep{Thorngren+(2016)} can be amassed from giant impacts \citep{GinzburgChiang(2020)}. Impactors and the momentum they impart to a growing planet excite a spectrum of seismic modes  (Fig.~\ref{fig:ModeVis}; Sec.~\ref{sec:ModeAmp}).  Once excited, normal mode oscillations can persist over timescales comparable to a young planet's age (Sec.~\ref{sec:ModeDamp}).  We have applied our calculations to $\beta$ Pic b, a $\sim$10-20 Myr old, 13 Jupiter mass planet still embedded in a debris disk. The giant planet's eccentricity of $\sim$0.15 and heavy metal content of $\sim$100-300 Earth masses \citep[e.g.][]{GRAVITY2020, Wang2025} suggest a violent dynamical upbringing including planet-scale collisions.  We showed that JWST photometry could detect the $\ell = 0$, $n=1$ $p$-mode excited in $\beta$ Pic b by a merger with a Neptune-mass planet, if such a collision occurred within the past $\sim$9--18 Myr (Sec.~\ref{sec:LumVar}).

Seismology offers a direct window into giant planet interiors.  
The longest-lived fundamental modes have frequencies comparable to the planet's dynamical frequency $\sim$$(G \Mp/\Rp^3)^{1/2}$, and would constrain the planet's bulk density. Pressure mode frequencies vary with the inverse of the sound-crossing time $\sim (\int c_{\rm s}^{-1} \der r)^{-1}$, and could measure the temperature of the planet's interior. If rotational splitting of a mode were detected, the planet's spin could be inferred \citep[e.g.][]{Huber+(2013), Ong(2025)}. 
High-order acoustic modes ($n>1$) could be used to constrain the interior sound speed and its gradient  \citep[e.g.][]{Aerts(2021)}.  The detection of low-frequency gravity modes could point to regions of stable stratification, as has been done for Saturn \citep[e.g.][]{Fuller(2014), MankovichFuller(2021)}.  

Our treatment of seismic oscillations and their detectability can be improved. We modeled the planet as a blackbody. Accounting for the wavelength-dependent opacity of the atmosphere may alter predicted flux variations in certain wavebands \citep[e.g.][]{Townsend(1997)}. We also considered only the linear behavior of normal modes. Non-linear effects could affect the mode spectrum, and enhance damping rates. 

Impacts are not the only way to excite oscillations in giant planets. Hot and warm Jupiters may form through high-eccentricity migration, a process whereby tidal gravitational forces from the host star excite the lowest-frequency fundamental mode to large amplitudes \citep[e.g.][]{Mardling(1995a), Mardling(1995b), IvanovPapaloizou(2004), VickLai(2018), Wu(2018), Vick+(2019), Teyssandier+(2019)}.  Because directly-imaged planets have large separations from their hosts, any oscillations detected in their light-curve will be excited by impacts, not tidal forcing ($\beta$ Pic b has a semi-major axis of $\sim$10 AU and eccentricity of $\sim$0.15).  However, for shorter separation, eccentric planets, such as HD 80606 b \citep{Sikora+(2024)} and HAT-P-2 b \citep{Jacobs+2025}, infrared light curves may exhibit variations from tidally-excited $f$-modes.

\vspace{0.2in}
\noindent 
This work was supported by a 51 Pegasi b Heising-Simons Fellowship awarded to JJZ. EC acknowledges support from NSF AST grant 2205500 and a Simons Investigator grant. YZ acknowledges support from STScI HST data analysis grants associated with programs GO-15830, GO-16036, and GO-17427.

\vspace{5mm}

\software{astropy \citep{Astropy_1,Astropy_2},  
          GYRE \citep{TownsendTeitler(2013), TownsendZweibel(2018), GoldsteinTownsend(2020)}
          MESA \citep{Paxton+(2011),Paxton+(2013),Paxton+(2015),Paxton+(2018),Paxton+(2019),Jermyn+(2023)}, 
          numpy \citep{numpy_cite},
          pandas \citep{pandas_cite},
          PyVista \citep{pyvista},
          scipy \citep{scipy_cite}
          }

\appendix

\section{\texttt{MESA} and \texttt{GYRE} inlists}
\label{app:inlist}

We modify the \texttt{MESA} example to calculate the evolutionary history of a brown dwarf.  Importantly, we relax the assumption that the radiative zones are sticky, which causes the effective temperature to become nonphysical when the radial grid resolution is increased.  We first create a planet/brown dwarf, and allow it to relax for $10^5$ years with low resolution, otherwise time-step issues are encountered.  Below is our \texttt{inlist\_BD\_start} inlist:
\begin{verbatim}
&star_job
  show_log_description_at_start = .false. 
  create_initial_model = .true.
  save_model_when_terminate = .true.
  save_model_filename = 'BD_start.mod'
  required_termination_code_string = 'max_age'
  mass_in_gm_for_create_initial_model = 2.46757d31
  ! 13 m_jupiter
  radius_in_cm_for_create_initial_model = 3.5d10
  ! 5 r_jupiter
/ 
&kap
  Zbase = 0.02
  kap_lowT_prefix = 'lowT_Freedman11'
/ 
&controls
  write_pulse_data_with_profile = .true.
  pulse_data_format = 'GYRE'
  add_atmosphere_to_pulse_data = .true.
  mlt_make_surface_no_mixing = .false.   
  convergence_ignore_equL_residuals = .true.
  use_gold2_tolerances = .true.
  initial_z = 0.02
  max_age = 1.d5
  energy_eqn_option = 'dedt'
  warning_limit_for_max_residual = 1d99   
  num_trace_history_values = 2
  trace_history_value_name(1) = 'rel_E_err'
  trace_history_value_name(2) = 'log_rel_run_E_err'
  max_resid_jump_limit = 1d15
  max_corr_jump_limit = 1d15
  photo_interval = 10
  profile_interval = 10
  max_num_profile_models=5000
  history_interval = 1
  terminal_interval = 1
  write_header_frequency = 10
/
\end{verbatim}
We then increase the spatial resolution after the planet exceeds the age of $10^5$ years. A new inlist \texttt{inlist\_BD\_highres} loads \texttt{BD\_start.mod} in \texttt{\&star\_job}, while the following lines in the \texttt{\&controls} namelist increases the number of radial grid-points:
\begin{verbatim}
  mesh_delta_coeff = 0.1
  max_allowed_nz = 8000000
  max_surface_cell_dq = 1.d-17
  min_dq = 1.d-18
  R_function_weight = 2
  R_function_param = 1.d-5
  R_function3_weight = 10000
  P_function_weight = 80
  when_to_stop_rtol = 1.d-6
  when_to_stop_atol = 1.d-6
\end{verbatim}
We then calculate the mode properties, including the quasi-adiabatic perturbations to the entropy and radiative luminosity, by running the \texttt{GYRE} code on one of the \texttt{MESA} models:
\begin{verbatim}
&model
  model_type = 'EVOL'
  file = 'LOGS_highres/profile19.data.GYRE'
  file_format = 'MESA'
/
&mode
  l = 1
/
&mode
  l = 2
/
...
&mode
  l = 10
/
&osc
  inner_bound = 'REGULAR'
  outer_bound = 'JCD'
  outer_bound_branch = 'E_NEG'
  quasiad_eigfuncs = .TRUE.
  nonadiabatic = .FALSE.
/
&num
  diff_scheme = 'MAGNUS_GL4'
  n_iter_max = 300
/
&scan
  grid_type = 'LINEAR'
  freq_min = 0.5
  freq_max = 30
  n_freq = 100
/
&grid
  w_osc = 0.5
  w_exp = 0.2
  w_ctr = 0.1
  w_str = 0.1
  resolve_ctr=.TRUE.
  dx_max = 0.00001
/
&ad_output
  summary_file = 'summary/summary_ad.h5'   
  summary_item_list = 
  'M_star,R_star,l,n_pg,omega,E_norm'
  detail_template = 
  'detail/detail_ad.l%l.n%n.h5'
  detail_item_list = 
  'l,n_pg,omega,x,xi_r,xi_h,eul_Phi,eul_rho,
  lag_rho,eul_P,lag_P,lag_L,lag_T,lag_S'
/
\end{verbatim}
Notably, we use the \texttt{JCD} outer boundary condition \citep{Christensen-Dalsgaard2008}, which matches the solution to an oscillation penetrating into an isothermal atmosphere, which decays its energy density with height.  See Appendix~\ref{app:isotherm} for details.

\section{Heat After Impact}
\label{app:heat}

\begin{figure}
\centering
\includegraphics[width=\linewidth]{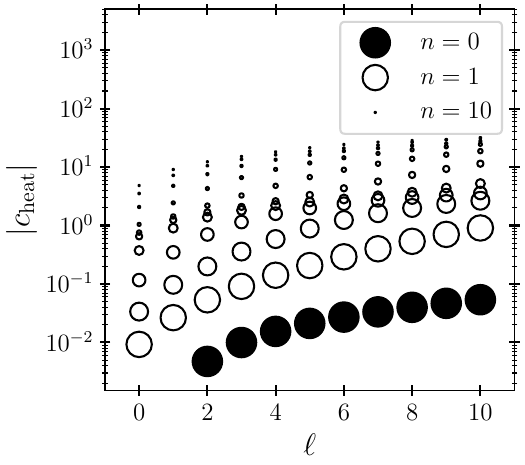}
\caption{Amplitudes due to work done by impactor heating $|c_{\rm heat}|$, for modes with radial node numbers $n$ and angular degrees $\ell$ indicated.  Symbol size decreases with $n$, from $n=0$ ($f$-modes, filled circles) to $n=10$ ($p$-modes, open circles).  We set $\eta = 0.5$ and $r_{\rm bub} = 0.995 \Rp$.  Notice $|c_{\rm heat}|$ is comparable to $|c_{\rm mom}|$ for the $f$-modes and $p$-modes considered here (Fig.~\ref{fig:amps}).
\label{fig:amp_heat}}
\end{figure}

In addition to the impact's momentum, heat deposited in the envelope of the planet can do work on normal modes \citep[e.g.][]{Lognonne+(1994), DombardBoughn(1995), WuLithwick(2019)}.  In this appendix, we estimate the amplitudes of oscillations after the impact heats the upper envelope.  We assume after impact at $t=0$, the impactor's heat causes oscillations to feel a constant pressure $P_{\rm bub}$ inside a bubble with a surface located at the radial locations $\br_{\rm bub}$ with a surface unit normal vector of $\hn_{\rm bub}$.  The pressure then exerts a force per unit area of
\begin{equation}
    \rho f_{\rm heat} = P_{\rm bub} \Theta(t) \delta(\br - \br_{\rm bub}) \hn_{\rm bub},
\end{equation}
where $\Theta(t)$ is the Heavyside step function, while $\delta(\br)$ is the one-dimensional Dirac delta function.  By using Gauss's law multiple times \citep[e.g.][]{DombardBoughn(1995)}, we can then express
\begin{align}
    &\langle \widetilde \bxi | {\bm f}_{\rm heat} \rangle = \int \widetilde \bxi^* \bcdot \left[ P_{\rm bub} \Theta(t) \delta(\br - \br_{\rm bub}) \hn_{\rm bub} \right] \rho \der V
    \nonumber \\
    &= P_{\rm bub} \Theta(t) \oint_{\rm bub} (\hn_{\rm bub} \bcdot \widetilde \bxi^*) \rho \der A
    \nonumber \\
    &= P_{\rm bub} \Theta(t) \int_{\rm bub} (\bdel \bcdot \widetilde \bxi^*) \rho \der V \nonumber \\
    &\simeq P_{\rm bub} V_{\rm bub} \Theta(t) (\bdel \bcdot \widetilde \bxi^*)_{\rm bub},
    \label{eq:overlap_bub}
\end{align}
where $V_{\rm bub}$ is the volume of the bubble, and $(\bdel \bcdot \widetilde \bxi^*)_{\rm bub}$ is evaluated at the center of the bubble.  

The work done on the oscillations, $P_{\rm bub} V_{\rm bub}$, should be some fraction $\eta$ of the impactor's energy $E_{\rm imp} = G M_{\rm p} m_{\rm imp}/R_{\rm p}$:
\begin{equation}
    P_{\rm bub} V_{\rm bub} = \eta E_{\rm imp} = \eta \frac{G M_{\rm p} m_{\rm imp}}{R_{\rm p}}.
    \label{eq:work_bub}
\end{equation}
Inserting equations~\eqref{eq:overlap_bub} and~\eqref{eq:work_bub} into equation~\eqref{eq:dot_c}, and also using $(\bdel \bcdot \widetilde \bxi^*) = -\Delta \widetilde \rho^*/\rho$, we find
\begin{equation}
    c_{\rm heat} = -\eta \frac{m_{\rm imp}}{M_{\rm p}} \left( \bdel \bcdot \widetilde \bxi^* \right)_{\rm bub} = -\eta \frac{m_{\rm imp}}{M_{\rm p}} \left( \frac{\Delta \widetilde \rho^*}{\rho} \right)_{r_{\rm bub}} \sqrt{ \frac{2\ell + 1}{4 \pi} },
\end{equation}
where we have assumed the bubble is located at the radius $r_{\rm bub}$, at the planet's pole.  We find $c_{\rm heat}$ to be comparable in magnitude to $c_{\rm mom}$ (compare Fig.~\ref{fig:amp_heat} to Fig.~\ref{fig:amps}).

\section{Heating at the Photosphere}
\label{app:isotherm}

Although many calculations take the Lagrangian pressure perturbation to vanish at the planet's photosphere ($\Dg \widetilde P/P = 0$ at optical depth $\tau = 2/3$), this boundary condition neglects how the oscillation affects the planet's atmosphere, and in particular how modal energy is lost to the atmosphere.  We follow instead Section 5.4 from the lecture notes available at \url{https://users-phys.au.dk/jcd/oscilnotes/Lecture_Notes_on_Stellar_Oscillations.pdf} (see also e.g. \citealt{GoldreichWu(1999)}) to calculate how the oscillation penetrates the planet's atmosphere.  We then use this result to calculate the work done by radiative diffusion at the photosphere.

We define the planet's photospheric radius $\Rp$ as the location where the optical depth $\tau = 2/3$, and take the layers above to be isothermal, so that the pressure scale height $H_P = -(\der \ln P/\der r)^{-1}$ and density scale height $H = (\der \ln \rho/\der r)^{-1}$ are equal:
\be
H_P = H = \frac{k_{\rm B} T}{g \mu m_{\rm H}} = \text{constant}.
\ee
Hydrostatic equilibrium
\be
\frac{\der P}{\der r} = -g \rho = - \frac{P}{H_P}
\ee
gives the solutions
\begin{equation}
    P(z) = P(\Rp) \e^{-z/H_P}, \hspace{5mm} \rho(z) = \rho(\Rp) \e^{-z/H_P},
\end{equation}
where $z = r - \Rp$ is the height above the surface, $k_{\rm B}$ is the Boltzman constant, $\mu$ is the mean molecular weight, and $m_{\rm H}$ is the mass of the hydrogen atom.  Because $H_P \ll \Rp$, we will also assume that $z \ll \Rp$ everywhere, so that the adiabatic index $\Gamma_1$, adiabatic sound-speed $c_{\rm s}$ and Brunt-V\"ais\"all\"a frequency $N$ satisfy
\begin{align}
    c_{\rm s}^2 &= \frac{\Gamma_1 P}{\rho} = \text{constant}, \\
    N^2 &= \frac{g}{H_P} \left( 1 - \frac{1}{\Gamma_1} \right) = \text{constant}.
\end{align}

Near the surface ($H_P/r \ll 1$), the equations describing the radial displacement $\widetilde \xi_r$ and Eulerian pressure perturbation $\dg \widetilde  P$ under the Cowling approximation ($\dg \widetilde \Phi \simeq 0$) are
\begin{align}
    &\frac{\der \widetilde \xi_r}{\der z} = \frac{1}{\Gamma_1 H_P} \widetilde \xi_r + \left( \frac{k_\perp^2}{\om^2} - \frac{1}{c_{\rm s}^2} \right) \frac{\dg \widetilde P}{\rho}, 
    \label{eq:dxirdz_isotherm} \\
    &\frac{\der}{\der z} \left( \frac{\dg \widetilde P}{\rho} \right) =  (\om^2 - N^2) \widetilde \xi_r + \frac{N^2}{g} \frac{\dg \widetilde P}{\rho}, \label{eq:ddgpdz_isotherm}
\end{align}
where $k_\perp = \sqrt{\ell(\ell+1)}/r$ is the horizontal wavenumber.  Because $k_\perp \ll H_P^{-1}$, we also make the plane-parallel approximation, and assume that radial variations are larger than those in the horizontal direction. Combining equations~\eqref{eq:dxirdz_isotherm}-\eqref{eq:ddgpdz_isotherm}, we have
\begin{equation}
    \frac{\der^2 \widetilde \xi_r}{\der z^2} - \frac{1}{H_P} \frac{\der \widetilde \xi_r}{\der z} + \frac{\om^2}{4 H_P^2 \om_{\rm ac}^2} \widetilde \xi_r \simeq 0,
    \label{eq:d2xirdr2_isotherm}
\end{equation}
where
\begin{equation}
    \om_{\rm ac} = \frac{c_{\rm s}}{2 H_P}
\end{equation}
is the acoustic cutoff frequency.  Assuming 
\begin{equation}
    \frac{\widetilde \xi_r}{r}, \frac{\Delta \widetilde P}{P} \propto \exp \left( \frac{\Lambda z}{H_P} \right),
\end{equation}
equation~\eqref{eq:d2xirdr2_isotherm} has a solution when
\begin{equation}
    \Lambda = \Lambda_\pm \equiv \frac{1}{2} \pm \frac{1}{2} \left( 1 - \frac{\om^2}{\om_{\rm ac}^2} \right)^{1/2}.\\
\end{equation}
The oscillation is evanescent ($\Lambda_\pm$ is real) for $\om < \om_{\rm ac}$, because pressure cannot provide a restoring force in the atmosphere.  The requirement that the energy density decays with height ($\rho \widetilde \xi_r^2 \to 0$ as $z \to \infty$) forces us to pick $\Lambda = \Lambda_-$.  For the $f$-modes and low-order $p$ modes of interest in this work, $\om \ll \om_{\rm ac} = 2\pi/(2.5 \ {\rm minutes})$ for our model, and
\begin{equation}
    \Lambda_- = \frac{1}{2} - \frac{1}{2} \left( 1 - \frac{\om^2}{\om_{\rm ac}^2} \right)^{1/2} \simeq \frac{\om^2}{4 \om_{\rm ac}^2}.
    \label{eq:Lambda_-}
\end{equation}

Within our gray, plane-parallel atmosphere, we assume radiation has no sources or sinks of energy, so the radiative flux streams freely ($F_{\rm rad} = \text{constant}$ when $z \ge 0$).  Since an Eddington atmosphere varies with optical depth as 
$T^4 = \frac{3}{4\sigma} F_{\rm rad}(\tau + \frac{2}{3})$
(with $\sigma$ the Stefan-Boltzmann constant), the Lagrangian perturbation to the radiative flux $\Dg F$ when the thermal time is long ($t_{\rm th} \om  \gg 1$) and the medium optically thin ($\tau \ll 1$) is given by
\begin{equation}
    \frac{\Delta F}{F_{\rm rad}} \simeq 4 \frac{\Dg T}{T} \simeq 4 \nabla_{\rm ad} \frac{\Dg P}{P},
    \label{eq:lagF_phot}
\end{equation}
where $\nabla_{\rm ad} = (\pd \ln T/\pd \ln P)_S$.  Therefore, the entropy perturbation at the photosphere is
\begin{equation}
    T \frac{\pd}{\pd t} \Dg S \bigg|_{r=\Rp} = - \frac{1}{\rho} \bdel \bcdot \Dg {\bm F} \bigg|_{r=\Rp} \simeq - \frac{4 \Lambda_- \nabla_{\rm ad} F_{\rm rad}}{\rho H_P} \frac{\Delta P}{P} \bigg|_{r=\Rp}.
    \label{DgS_phot}
\end{equation}
Also using the thermodynamic relation
\begin{align}
    &\frac{1}{\Gamma_1} \frac{\Dg P}{P} - \upsilon_T \frac{\Dg S}{C_P} = \frac{\Dg \rho}{\rho} = -(\bdel \bcdot \bxi) \simeq - \frac{\pd}{\pd z} \xi_r,
    \label{eq:divxi_thermo}
\end{align}
where $\upsilon_T = -(\pd \ln \rho/\pd \ln T)_S$, we can write $\widetilde \xi_r$ in terms of $\Dg \widetilde S$ and $\Dg \widetilde P$, which gives the leading-order lag between $\widetilde \xi_r$ and $\Dg \widetilde P^*$ at the planet's surface:
\begin{equation}
    \widetilde \xi_r = \left( \im \frac{4 \upsilon_T \nabla_{\rm ad} F_{\rm rad}}{\omega \rho C_P T} + \frac{H_P}{\Lambda_- \Gamma_1} \right) \frac{\Dg \widetilde P}{P}.
    \label{eq:xir_photo}
\end{equation}
Inserting~\eqref{eq:xir_photo} into equation~\eqref{eq:gamma_surf}, we arrive at equation~\eqref{eq:gamma_photo}.  

\bibliography{main}{}
\bibliographystyle{aasjournal}



\end{document}